\documentclass[twocolumngrid,prd,superscriptsize,reprint]{revtex4-1}
\pdfoutput=1
\usepackage[utf8]{inputenc} 
\usepackage{graphicx,xcolor,overpic,mathtools}
\usepackage{amsthm,amsmath,amssymb}
\usepackage[colorlinks]{hyperref}
\usepackage{textcomp}
\definecolor{coolblack}{rgb}{0.0, 0.18, 0.39}
\hypersetup{
    colorlinks = false,
   }
\PassOptionsToPackage{normalem}{ulem}
\usepackage{ulem} 
\usepackage{sidenotes}
\usepackage{bm}
\usepackage{url}
\usepackage{physics}
\usepackage[makeroom]{cancel}
\usepackage{cleveref}
\usepackage{tensor}
\usepackage{mathrsfs}
\usepackage{slashed}
\usepackage[mode=buildnew]{standalone}
\newcommand{\lag}{\mathcal{L}}

\newcommand{\Lt}{\widetilde{L}}
\newcommand{\comment}[1]{}
\newcommand{\Et}{\widetilde{E}}

\usepackage{pgfplots}

\usepackage{diagbox}

\NewDocumentCommand{\evat}{sO{\bigg}mm}{%
  \IfBooleanTF{#1}
   {\mleft. #3 \mright|_{#4}}
   {#3#2|_{#4}}%
}
\usepackage{enumerate}
\definecolor{azure}{rgb}{0.0, 0.5, 1.0}

\begin{document}

\title[]{Dense matter equation of state and phase transitions from \\ a Generalized Skyrme model}

\author{Christoph Adam}
\author{Alberto Garc\'ia Mart\'in-Caro}
\author{Miguel Huidobro}%
\author{Ricardo V\'azquez}
\affiliation{%
Departamento de F\'isica de Part\'iculas, Universidad de Santiago de Compostela and Instituto
Galego de F\'isica de Altas Enerxias (IGFAE) E-15782 Santiago de Compostela, Spain
}%
\author{Andrzej Wereszczynski}
\affiliation{
Institute of Physics, Jagiellonian University, Lojasiewicza 11, Krak\'ow, Poland
}%

\date[ Date: ]{\today}
\begin{abstract}
Skyrmion crystals are the field configurations which minimize the energy per baryon in the infinitely large topological charge sector of the Skyrme model, at least for sufficiently high density. They are, therefore, an important tool to describe the ground state of cold, symmetric nuclear matter at high density regimes. In this work, we analyze different crystalline phases and the existence of phase transitions between them within the generalized Skyrme model, with the ultimate goal of describing symmetric nuclear matter in a wide regime of densities. Furthermore, we propose a new energy-minimizing phase for densities lower than the nuclear saturation point ($n_0$) which also presents a good qualitative behavior in the zero density limit, thereby improving the description of strongly interacting matter in the region $n<n_0$. 
\end{abstract}
\maketitle
%\begin{minipage}{\textwidth}
\tableofcontents
%\end{minipage}

\section{Introduction}
Quantum Chromodynamics (QCD) is the non-abelian gauge theory which describes strongly interacting matter in terms of its fundamental degrees of freedom, namely quark and gluon fields. Despite the great success of QCD at high energy scales, in which it becomes a weakly interacting theory, its non-perturbative character at the low energy scale makes calculations of nucleons and nuclear matter properties extremely difficult, and other alternatives to the usual perturbative approach must be considered, like lattice QCD or phenomenological nuclear physics models, such as the Skyrme model \cite{skyrme1994non}.

Similarly to other effective approaches to strongly interacting matter, like chiral perturbation theory (ChPT) of relativistic mean-field theory (RMF), the basic fields of the Skyrme model are given by the fields that provide the physical, asymptotic particle states, concretely the meson fields. At variance with ChPT, however, a chiral expansion in powers of the typical momentum or energy of a physical process is not assumed. Instead, terms in the lagrangian with different scaling dimensions are treated on an equal footing, allowing for a balance between oppositely scaling terms and evading the Derrick theorem \cite{Derrick:1964ww}.
As a consequence of this balance, it is sufficient to introduce the mesonic fields as the basic degrees of freedom (DoF), because the baryons emerge as collective excitations or topological solitons ("skyrmions") from the nonlinear interactions of the mesons \cite{skyrme1962unified}. 
The Skyrme model was, in fact, originally proposed by T. Skyrme in 1961 precisely with the aim of describing baryons within a self-interacting pion field theory. 
The Skyrme model differs in this respect from both ChPT and RMF, where the baryons must be introduced as independent DoF. Furthermore, both the conservation of baryon number - which is identified with the topological charge of the skyrmions - and the extended character of baryons are built-in properties of the Skyrme model. 

Later, it was shown that QCD in the large $N_c$ (the number of colors) limit reduces to a weakly interacting mesonic field theory in which baryons share the properties of topological solitons  \cite{tHooft:1973alw, witten1979current}. Independent support for the Skyrme model is provided by its derivation from holographic QCD, both for the original \cite{Sakai:2004cn,Sakai:2005yt} and for the generalized Skyrme model \cite{Bartolini:2017sxi}.

First attempts of reproducing the properties of nucleons and nuclei within the Skyrme model were partially successful, but with a typical precision of only about $30\%$.  In addition, there remained some relevant discrepancies, like the too large nuclear binding energies and the shell-like matter distribution of the Skyrme model solutions \cite{manton2004topological}. Recent results, however, demonstrate that both a quantization procedure beyond the rigid rotor approximation and the addition of new terms in the Skyrme lagrangian can solve many of these problems \cite{Adam_2010, adam2010bps, Gillard:2015eia, Gudnason:2016mms, Naya:2018kyi} and lead to much more precise results. Concretely, generalized Skyrme models which lead to realistic binding energies are discussed, e.g., in \cite{Adam_2010, adam2010bps, Gillard:2015eia, Gudnason:2016mms}, whereas in \cite{Naya:2018kyi} it is demonstrated that the inclusion of the rho meson allows to find the known cluster structures of light nuclei. Finally, in \cite{Lau:2014baa} and \cite{Halcrow:2016spb}, the excitation spectra of carbon-12 and oxigen-16 are reproduced with an astonishing precision where, in the latter case, the quantization of both rotational and vibrational DoF has been taken into account. There has also been important progress in the Skyrme model description of the nucleon-nucleon forces \cite{Halcrow:2020gbm}. 

On the other hand, neutron stars (NS) have become our most useful resource for studying the behavior of nuclear matter at ultra-high densities \cite{Lattimer:2004pg}. Indeed, with the advent of gravitational wave astronomy, a deeper insight in the Equation of State (EoS) of strongly interacting matter has been provided by recent gravitational wave events involving binary neutron star mergers \cite{Abbott_2017}. Despite the large theoretical and observational effort employed in the last decades, the EoS of nuclear matter at a range of densities much higher than the nuclear saturation density is still not fully understood. From all the different approaches to the study of dense nuclear matter, the Skyrme model stands out as a relatively simple effective model with a low number of free parameters. Moreover, an equation of state based on this model (and its generalization) has been recently proposed in \cite{Adam:2020yfv} and shown to yield reasonable results in predicting the properties of neutron stars such as the mass-radius relations or  the quasi-universal relations between the moment of inertia, the deformability and the quadrupolar moment of slowly rotating and tidally deformed stars \cite{naya2019neutron,Adam:2020aza}.

However, the EoS proposed in \cite{Adam:2020yfv} was based on the interpolation between two different submodels of the general Skyrme model, namely, the standard Skyrme model, that predicts a crystalline state of the dense nuclear matter, and the BPS submodel, in which matter behaves as a perfect fluid \cite{Adam:2014nba}. The transition between the crystal and fluid phases was modeled as a smooth crossover, where an additional free parameter describing the point at which the transition takes place had to be introduced.

Moreover, within the Skyrme model literature it has been established that the configurations that minimize the energy per baryon at large baryon number correspond to crystalline solutions, in which skyrmions (or half-skyrmions, see sect. III) are arranged in a periodic fashion respecting some particular (discrete) symmetries. Indeed, configurations with different symmetries and energies have been proposed in order to find the one with minimal energy. However, some symmetries could be more energetically favourable than others at different density regimes. This is indeed what was found for the standard Skyrme model in \cite{klebanov1985nuclear}, in which the existence of a phase transition between different crystalline configurations is predicted at a certain density.

In the present paper, our goal is to construct different solutions both of crystalline and non-crystalline types  of the full generalized Skyrme model and to study their behavior
for a wide range of densities. The main aim is to determine which configurations minimize the energy per baryon in the different regimes, and to find the corresponding equation of state (EoS). The resulting classical skyrmionic matter and its EoS should provide an interesting starting point for the description of strongly interacting matter. For a completely realistic description, however, most likely further modifications like quantum corrections or the effects of additional fields have to be taken into account. 

This paper is organised as follows: In the second section we introduce the general Skyrme model, from which we will construct the crystalline solutions. In the third section we review the procedure of how to construct crystal-like solutions following \cite{kugler1989skyrmion}. In section IV, we study the problem of the inhomogeneous phase for nuclear matter at intermediate densities, in which the Skyrmion crystal ceases to be a good ansatz for the field, as the energy per baryon starts to grow. Then we use these solutions to obtain an equation of state (EoS) for classical skyrmionic matter. Finally we  end with some conclusions and possible future directions. We always assume units such that the speed of light $c=1$. Further, we use the mostly minus metric convention.

\section{Generalized Skyrme Lagrangian}
The general Skyrme model is described by the following lagrangian density,
\begin{align}
    \notag \lag = &\lag_{\rm Sk} + \lag_{\rm BPS} = \left( \lag_2 + \lag_4 \right) + \left( \lag_6 + \lag_0 \right)= \\[2mm]
    \notag= -\frac{f^2_{\pi}}{16}&\Tr\left\{L_{\mu}L^{\mu}\right\} + \frac{1}{32e^2}\Tr\left\{\left[L_{\mu},L_{\nu}\right]^2\right\} \\[2mm] 
    &- \lambda^2 \pi^4\mathcal{B}_{\mu}\mathcal{B}^{\mu} + \frac{m^2_{\pi} f^2_{\pi}}{8}\Tr\left\{ U - I \right\}.
    \label{Lag}
\end{align}
Apart from the specific choice for the potential term $\mathcal{L}_0$ -  the pion mass term - which could be replaced by a more general expression, the above lagrangian density is the most general one in terms of the pion field only which is both Poincare invariant and at most quadratic in time derivatives, such that a standard hamiltonian can be defined.  

We find it useful to regroup the full generalized model $\mathcal{L}$ into the standard part $\mathcal{L}_{\rm Sk} $ and the BPS part $\mathcal{L}_{\rm BPS}$, because some solutions of these submodels for large baryon number $B$ are relatively simple and have been widely studied, which will allow us to compare our full solutions to these limiting cases.
The second part $\mathcal{L}_{\rm BPS}$ is a BPS model, \textit{i.e.}, it has solutions saturating the corresponding Bogomol'nyi energy bound \cite{Adam_2010}. The lagrangian has 3 free parameters ($f_{\pi}, e, \lambda^2$) which will be used to fit the ground state of the solutions. The pion mass is set to its physical value $m_{\pi} = 140$ MeV. This model represents the dynamics of a $SU(2)$ field $U$, which always appears in the Lagrangian in terms of the Maurer-Cartan left-invariant current $L_{\mu} = U^{\dagger}\partial_{\mu}U$, except for the potential term $\mathcal{L}_0$. Static configurations of the field $U$ constitute maps from $\mathbb{R}^3$ to the the target space manifold, which can be identified with the three sphere $S^3$. 

For usual solitonic configurations, the requirement of finite energy implies that the field must take values in the vacuum manifold at spatial infinity, which, due to the potential term, corresponds to $U(\abs{x}\rightarrow \infty) = I$. This boundary condition, in turn, implies that finite energy configurations correspond to mappings from one-point compactified real space, $\mathbb{R}^3\cup \{\infty \} \sim S^3$ to $S^3$. These maps are classified by the third homotopy group of $S^3$, $\pi_3(S^3) = \mathbb{Z}$, so they can be labelled by an integer. Hence, the Skyrme model allows for topological soliton solutions, called Skyrmions, carrying an integer valued charge. This integer, the so-called topological degree, is identified with the baryon number $B$, and can be calculated as an integral of the topologically conserved current $\mathcal{B}^\mu$:
\begin{equation}
    B = \int d^3 \mathcal{B}^0, \hspace{2mm} \mathcal{B}^{\mu} = \frac{1}{24\pi^2} \epsilon^{\mu\nu\alpha\beta}\Tr\left\{ L_{\nu}L_{\alpha}L_{\beta} \right\},
    \label{TopoNumber}
\end{equation}
which is the same expressions that appears in the sextic term ($\mathcal{L}_6$) of the lagrangian \eqref{Lag}.

Solutions of the standard Skyrme model in the $B = 1$ sector with \cite{adkins1984skyrme} and without pion mass term \cite{adkins1983static}, and including the contributions from the zero mode quantization, have been found to reproduce the nucleon properties reasonably well. Later, these calculations were extended for higher values of $B$ \cite{battye2009light} within the rational map approximation. This has also be done in the BPS model \cite{Adam:2013wya, Adam:2013tda}, obtaining quite accurate results according to experimental data, and with no approximation since the symmetries of the BPS model allows an analytical treatment of the solutions. The generalized Skyrme lagrangian \eqref{Lag} was used to reproduce nucleons, as well \cite{Ding:2007xi}. These last results will be compared to the ones we obtain from the condition to reproduce infinite nuclear matter.

To describe NS, on the other hand, we need to find solutions for $B$ of the order of $B \sim 10^{57}$, the number of baryons in the Sun. Then, we should think about how skyrmions arrange under these conditions. It was Klebanov \cite{klebanov1985nuclear} who proposed a kind of crystalline solution with the aim of describing the highly compressed interiors of neutron stars. As usually, when considering crystalline configurations, we will define a unit cell for each symmetry and work with it. Hence, in order to describe these solutions, we may define the Skyrme fields as mappings from the finite size unit cell (which has finite energy) to the target manifold.

We would like to remark that, although the boundary conditions imposed on the Skyrme fields are different from those of regular solitonic solutions, the topological properties of the field configurations remain the same. Indeed, a cubic lattice with periodic boundary conditions is mathematically equivalent to a three torus, $T^3$, so that crystalline configurations are described by maps $U_{\rm crystal}:T^3\rightarrow S^3$. As $T^3$ is still a compact and oriented manifold, mappings from $T^3$ to $S^3$ are still characterized by their topological degree, as ensured by Hopf's degree theorem \footnote{In fact, the theorem asserts that the topological degree is the \emph{only} homotopy invariant in such situations}.

Then, in the standard Skyrme model the solution that minimizes the energy is a crystalline configuration. On the other hand, we know that the BPS model solutions behave like a perfect fluid, due to the symmetry under volume preserving diffeomorphisms of $\mathcal{L}_{\rm BPS}$ (in fact, one can exactly identify the field configurations in the BPS submodel  $\mathcal{L}_{\rm BPS}$ with a perfect fluid, as can be shown from a careful analysis of the corresponding stress-energy tensor \cite{Adam_2010}). However, since the sextic term is only important at high densities \cite{Adam:2020yfv}, we expect that the crystal solution is still the ground state solution in the generalized Skyrme lagrangian.

In order to construct Skyrme crystal solutions it is useful to define dimensionless units of length and energy ($\vec{r} = (x,y,z)^{\rm t}$)
\begin{equation}
    \vec{r} = \frac{1}{f_{\pi}e}\vec{\widetilde{r}}, \hspace{3mm} E = \frac{3\pi^2f_{\pi}}{e}\Et.
\end{equation}
These units are frequently used in the Skyrme model, so they are useful to compare the results. It is commonly known that the energy of the $B = 1$ skyrmion in the standard Skyrme model is $\Et = 1.23$, whereas the topological (Skyrme-Faddeev) bound \cite{skyrme1962unified,Faddeev:1976pg} on the energy reads $\Et \ge 1$ in these units.

The field $U$ can be parametrized as an expansion in the $SU(2)$ Lie algebra generators:
\begin{equation}
    U = \sigma + i \pi_k \tau_k,
    \label{Ufield}
\end{equation}
where the $\pi_k$ ($k$ = 1, 2, 3) represent the pions, $\tau_k$ are the Pauli matrices, and the fields satisfy the unitarity condition $\sigma^2 + \pi_i\pi_i = 1$. We will work with static solutions $\partial_0 U = 0$, then the energy is simply $E = -\int d^3x \lag$. Inserting \eqref{Ufield} in \eqref{Lag} and \eqref{TopoNumber} we can calculate the energy and the baryon number:
\begin{align}
    \notag\Et = \frac{1}{24\pi^2}&\int d^3\widetilde{x} \left[ -\frac{1}{2}\Tr\left\{L_iL_i\right\} - \frac{1}{4}\Tr\left\{\left[L_i,L_j\right]^2\right\} + \right.\\[2mm]
    \notag&\left. 8\lambda^2 \pi^4 f^2_{\pi}e^4 \mathcal{B}^0\mathcal{B}^0 + \frac{m^2_{\pi}}{f^2_{\pi}e^2}\Tr\left\{ I - U \right\}  \right] =\\[2mm]
    \notag=\frac{1}{24\pi^2}&\int d^3\widetilde{x} \left[ \partial_i n_a\partial_i n_a + \left( \partial_i n_a \partial_j n_b - \partial_i n_b\partial_j n_a \right)^2 + \right. \\[2mm]
    &\left. C_6\left( \epsilon_{abcd}n_a\partial_1 n_b\partial_2 n_c \partial_3 n_d \right)^2 + C_0\left( 1-\sigma \right)\right], \label{Energy} \\[2mm]
    B = -\frac{1}{2\pi^2}&\int d^3\widetilde{x} \:\epsilon_{abcd}n_a\partial_1 n_b\partial_2 n_c \partial_3 n_d,
    \label{Baryon}
\end{align}
where we have define the unit vector $n_a = (\sigma, \pi_i)$, and the constants $C_6 = 2\lambda^2f^2_{\pi}e^4$, $C_0 = \frac{2m^2_{\pi}}{f^2_{\pi}e^2}$.

\section{Crystal solutions in the Skyrme model}
Two $B=1$ skyrmions are in the maximally attractive channel if the second skyrmion is isorotated by $\pi$ relatively to the first one, about an axis perpendicular to the distance vector between the two skyrmions. It can be checked easily that the maximally attractive orientation of skyrmions can be extended to a cubic arrangement, such that all skyrmions forming the cubic lattice are maximally attracted by all their nearest neighbours. This led Klebanov \cite{klebanov1985nuclear} to consider a Skyrme crystal based on an infinite periodic lattice with cubic symmetry. At low densities, the solution is described by spherically symmetric skyrmions located in the corners of the cube. The fact that nearest neighbours must be mutually isorotated to be in the maximally attractive channel translates into a particular set of symmetries for the field in the unit cell (simple cubic and periodic) that must be imposed. Then, the solution is found by minimizing the static energy functional \eqref{Energy}.

Most Skyrme crystal calculations have been performed for the standard Skyrme model $\mathcal{L}_{\rm Sk}$. We shall, therefore, briefly review these crystals and their symmetries before presenting our own results. In all cases, skyrmions (or half-skyrmions) are located at the vertices of a cubic lattice, and we call the distance between two nearest neighbours $L$ (or $\Lt$ in dimensionless units).  The Skyrme fields, however, are {\em not} periodic under a lattice translation by $L$, because of the necessity to isorotate nearest neighbours. They are, however, periodic for $2L$ translations. The unit cell of the crystal is, therefore, a cube with sidelength $2L$ (or $2\Lt$) in all cases.

The total energy of a crystal is infinite since it is, by construction, infinitely extended. Nevertheless, the energy per baryon number (here $\Et_{\text{cell}}$ is the dimensionless energy of the unit cell, and $N_{\text{cells}}$ is its baryon number),
\begin{equation}
    \frac{\Et}{B} = \frac{N_{\text{cells}}\:\Et_{\text{cell}}}{N_{\text{cells}}\:B_{\text{cell}}},
\end{equation}
remains finite. Then we can work with a single unit cell, which has finite energy, and calculate its baryon number and energy. The unit cell is characterized by the sidelength $2\Lt$, whereas its energy changes for different values of $\Lt$. The curve $\Et(\Lt)$ is always found to have a minimum $E_{\rm min} = E(L_{\rm min})$ for a certain finite $L_{\rm min}$, which for the case of the cubic symmetry of  \cite{klebanov1985nuclear} takes the value $\Et_{\rm min}/B = 1.08$. This value is only an $8\%$ higher than the Skyrme-Faddeev bound, 
which indicates that a crystalline arrangement of skyrmions is probably the field configuration with lowest energy (per baryon) for an infinite baryon number.

Different symmetries were computed to get closer to the (unattainable) Skyrme-Faddeev bound. In \cite{goldhaber1987maximal}, Manton and Goldhaber proposed an additional symmetry to the Klebanov crystal, motivated by the dynamics of the two-skyrmion configuration. This introduced a new solution based on half-skyrmions, which can be thought of as a body-centered cubic (BCC) arrangement, in which one half-skyrmion solution is located in the center of the cube and the other half-skyrmions in the corners.

Finally, in two different but almost simultaneous papers \cite{castillejo1989dense, kugler1989skyrmion} a new phase was proposed. They computed a crystal with face-centered cubic (FCC) symmetry of half-skyrmions using two different approaches. The resulting crystal is believed to be the crystal of lowest energy and is a good candidate for the ground state of the standard Skyrme model for infinite baryon number, with a minimal energy $E_{\rm min}$ per baryon which is only $3.8\%$ above the energy bound.

The influence of the sextic and potential terms in the crystalline phases of the Skyrme model was already investigated in \cite{perapechka2017crystal} from a more formal point of view. In our paper, we extend these studies, considering physical values of the parameters of the Skyrme model and focusing on the extraction of an equation of state for the ground state of symmetric nuclear matter.

All the crystals mentioned above have the cubic symmetries in common. They are given by the following combined transformations,
\begin{align}
    \notag\text{A}_1&: (x,y,z) \rightarrow (-x,y,z), \\[2mm] &(\sigma,\pi_1,\pi_2,\pi_3) \rightarrow (\sigma,-\pi_1,\pi_2,\pi_3), \label{A1}\\[2mm]
    \notag\text{A}_2&: (x,y,z) \rightarrow (y,z,x), \\[2mm] &(\sigma,\pi_1,\pi_2,\pi_3) \rightarrow (\sigma,\pi_2,\pi_3,\pi_1).
    \label{A2}
\end{align}

The simple cubic (SC) crystal of Klebanov has an additional periodicity symmetry,
\begin{align}
    \notag \text{A}_3&: (x,y,z) \rightarrow (x+L,y,z), \\[2mm] &(\sigma,\pi_1,\pi_2,\pi_3) \rightarrow (\sigma,-\pi_1,\pi_2,-\pi_3).
\end{align}
This symmetry locates the center of the skyrmions in the corners of the cube, isorotated with respect to their nearest neighbours. Owing to the translational invariance of A$_3$, the energy and baryon densities are periodical in $L$. Since each skyrmion contributes 1/8 to the baryon number and the cube has 8 corners, the baryon number of this cube is 1. However, the fields are periodical in $2L$ (as follows from the symmetry A$_3$), and the unit cell is a cube of length $2L$.

\begin{figure}[h!]
    \centering
    %\hspace{-1.4cm}
    \includegraphics[scale=0.3]{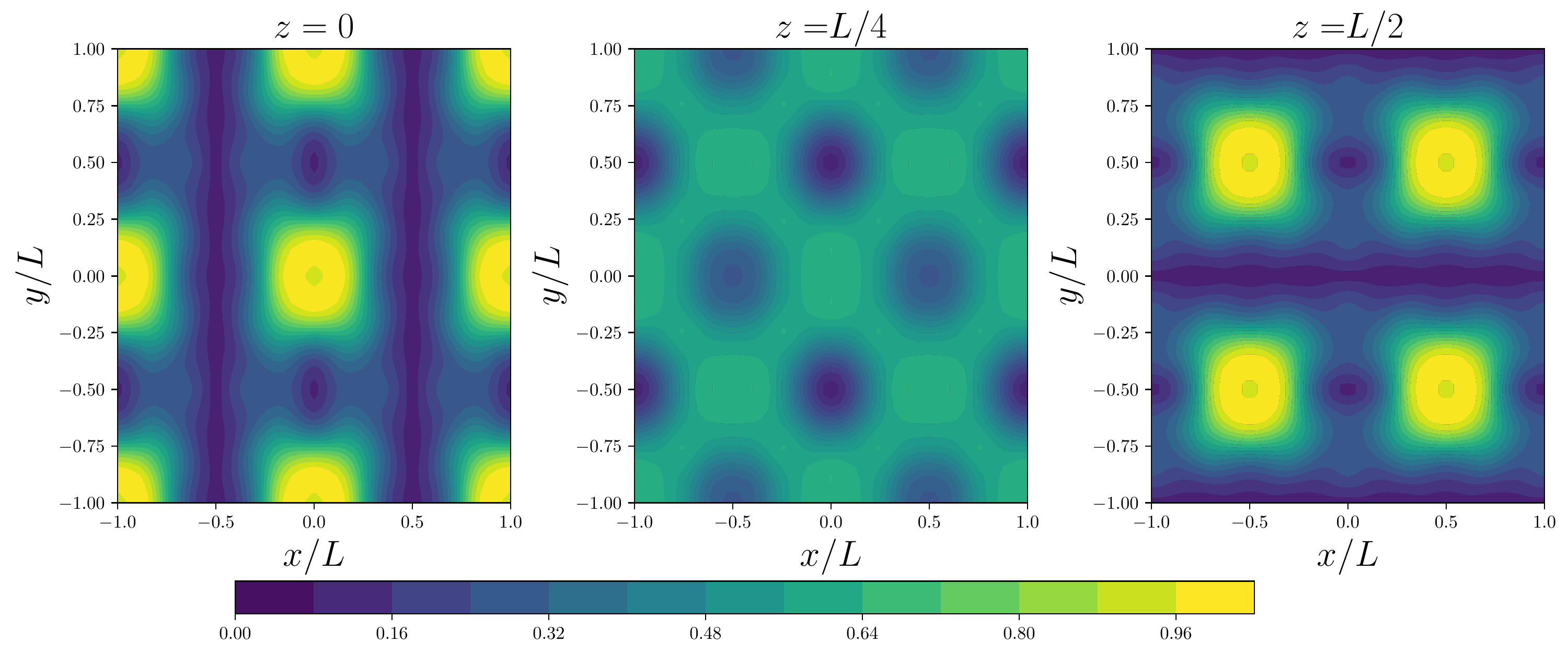}
    \caption{\small Energy contour plots for the unit cell of the body-centered-cubic (BCC) crystal. The plots show energy density surfaces for different heights within the unit cell.}
    \label{Figure.B}
\end{figure}

The BCC half-skyrmion phase shares the same symmetries of the SC phase ($\text{A}_1, \text{A}_2, \text{A}_3$), plus one additional symmetry,
\begin{align}
    \notag\text{B}_4&: (x,y,z) \rightarrow (L/2-z,L/2-y,L/2-x), \\[2mm] &(\sigma,\pi_1,\pi_2,\pi_3) \rightarrow (-\sigma,\pi_2,\pi_1,\pi_3).
\end{align}
In this phase, the cube of length $L$ has in its center a half skyrmion (with $\sigma = -1$ at the center) and 8 other half skyrmions (with $\sigma = +1$) in the corners. An interesting result obtained from this half-skyrmion symmetry is that the mean value of $\sigma$ over the unit cell, denoted by $\langle \sigma \rangle$, vanishes identically. From this property it is obvious that the potential term in the lagrangian will exactly scale as $L^3$ in this phase (remember that $\mathcal{L}_0 \sim \sigma -1$). Again, the unit cell has length $2L$, therefore the integrals \eqref{Energy} and \eqref{Baryon} are perfomed between $-L$ and $L$. It is easy to deduce that the baryon number of the unit cell is 8. A typical energy density plot is shown in \cref{Figure.B}, where blue regions correspond to low density and yellow regions to high density.

The FCC symmetry of single skyrmions has two different symmetries besides A$_1$ and A$_2$,
\begin{align}
    \notag\text{C}_3&: (x,y,z) \rightarrow (x,z,-y), \\[2mm] &(\sigma,\pi_1,\pi_2,\pi_3) \rightarrow (\sigma,-\pi_1,\pi_3,-\pi_2), \\[2mm]
    \notag\text{C}_4&: (x,y,z) \rightarrow (x+L,y+L,z), \\[2mm] &(\sigma,\pi_1,\pi_2,\pi_3) \rightarrow (\sigma,-\pi_1,-\pi_2,\pi_3).
\end{align}

\begin{figure}[h!]
    \centering
    %\hspace{-1cm}
    \includegraphics[scale=0.3]{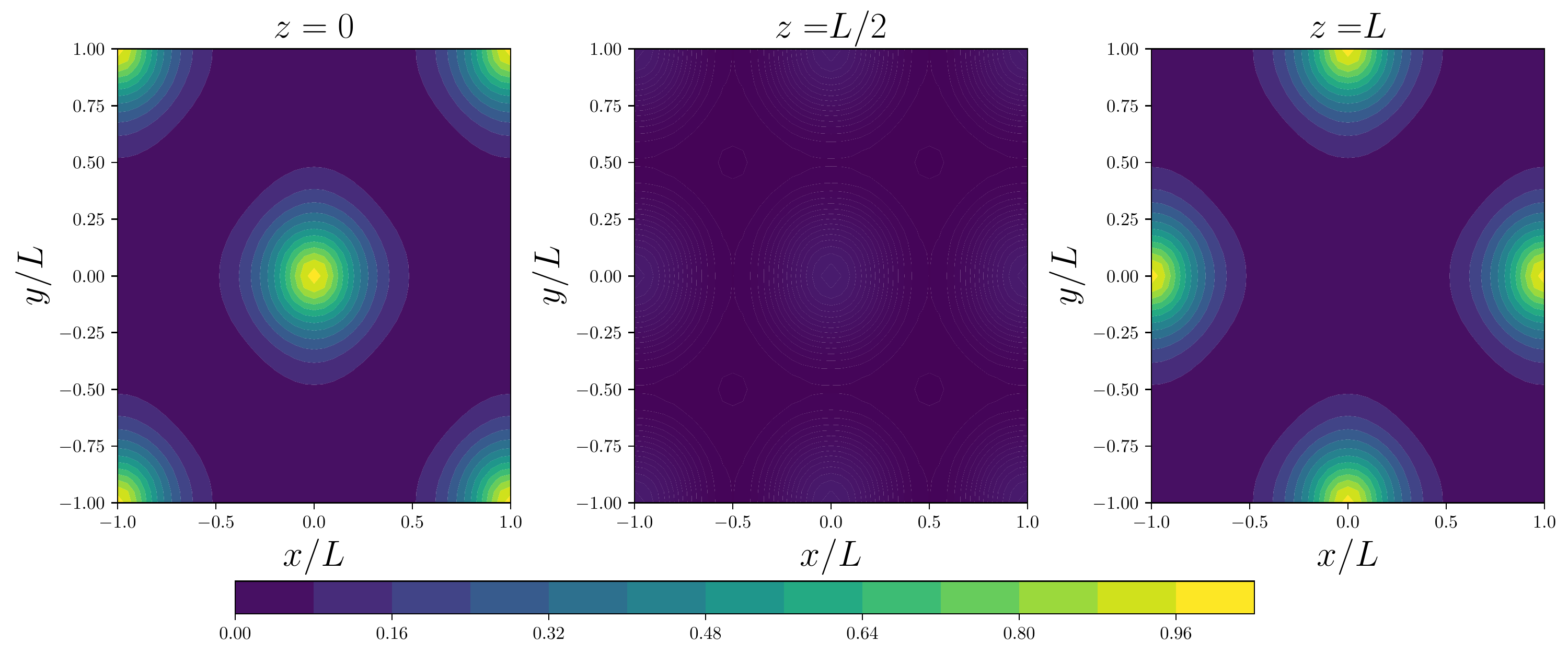}
    \caption{\small Energy contour plots for the unit cell of the face-centered cubic (FCC) crystal of skyrmions. The plots show energy density surfaces for different heights within the unit cell. Here we choose $z=0, L/2, L$ for the heights because in this case also the energy density has a $2L$ periodicity.}
    \label{Figure.C}
\end{figure}

The energy, baryon number and the fields are periodical in $2L$ in this case, a typical plot is shown in \cref{Figure.C}.

The FCC half-skyrmion symmetry shares $\text{C}_3$ and additionally:
\begin{align}
    \notag\text{D}_4&: (x,y,z) \rightarrow (x+L,y,z), \\[2mm] &(\sigma,\pi_1,\pi_2,\pi_3) \rightarrow (-\sigma,-\pi_1,\pi_2,\pi_3).
\end{align}

\begin{figure}[h!]
    \centering
    %\hspace{-1cm}
    \includegraphics[scale=0.3]{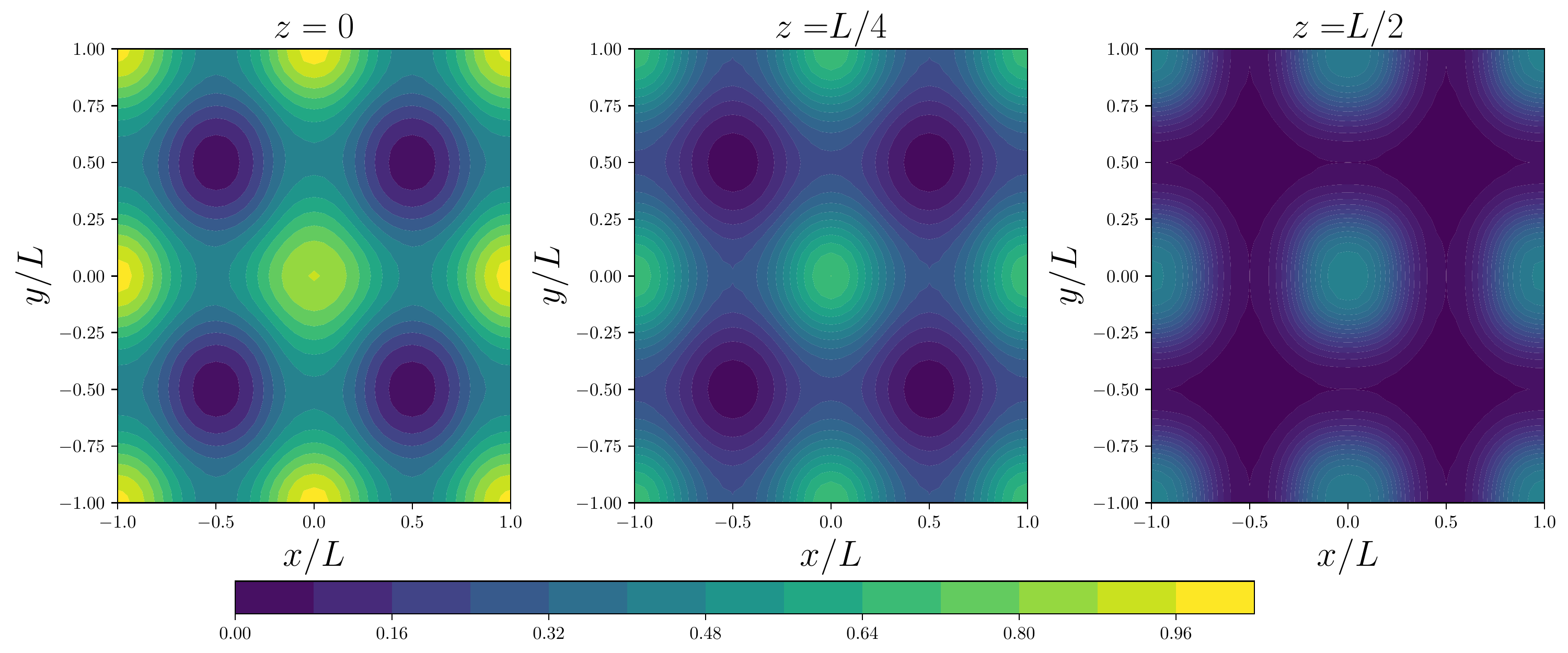}
    \caption{\small Energy contour plots for the unit cell of the face-centered cubic crystal of half-skyrmions (FCC$_{^1\!/_2}$). The plots show energy density surfaces for different heights within the unit cell.}
    \label{Figure.D}
\end{figure}

We can recover the FCC phase symmetry C$_4$ applying two D$_4$ transformations. In this phase, the energy and baryon number are periodic in $L$, but the fields have period $2L$, and it has, in fact, the appearance of a simple cubic phase of half-skyrmions, see \cref{Figure.D}. The baryon number per unit cell $(2L)^3$ is $B_{\rm cell} =4$ for both FCC crystals. Further, as in the BCC phase, $\langle \sigma \rangle$ vanishes, so the potential term contribution is already known and scales like $L^3$, like in the BCC phase.

From now on we will refer to the FCC phase of half-skyrmions as FCC$_{^1\!/_2}$.

Following \cite{kugler1989skyrmion}, we find that the cubic symmetries \eqref{A1} motivate the following Fourier-like expansion of the fields,
\begin{align}
    &\overline{\sigma} = \sum^{\infty}_{a,b,c = 0} \beta_{abc}\cos\left( \frac{a\pi x}{L} \right) \cos\left( \frac{b\pi y}{L} \right) \cos\left( \frac{c\pi z}{L} \right) \label{Sigmaexpan}\\[2mm]
    &\overline{\pi}_1 = \sum^{\infty}_{h,k,l = 0} \alpha_{hkl} \sin\left( \frac{h\pi x}{L} \right) \cos\left( \frac{k\pi y}{L} \right) \cos\left( \frac{l\pi z}{L} \right). \label{Piexpan}
\end{align}
Then, the fields $\pi_2$ and $\pi_3$ can be constructed applying the transformation A$_2$ on $\pi_1$. The bars over the fields denote that these fields do not satisfy the $SU(2)$ condition, hence we have to normalize them,
\begin{equation}
    n_a = \frac{1}{\sqrt{\overline{\sigma}^2 + \overline{\pi}_k\overline{\pi}_k}}\left( \overline{\sigma}, \overline{\pi}_k \right).
    \label{vector}
\end{equation}

Once the particular symmetries of a crystalline ansatz have been specified, the problem is reduced to a finite-dimensional optimization problem for the coefficients $\beta_{abc}$ and $\alpha_{hkl}$, which must be adequately chosen in order to minimize the energy \eqref{Energy} of the solution. Furthermore the symmetry properties associated to each phase can be used to reduce the number of independent parameters, since they result in some constraints between the coefficients. 

In the BCC phase, the following coefficients $\beta_{abc}$ and $\alpha_{hkl}$ may be nonzero
\begin{itemize}
    \item $h$, $k$ are odd, $l$ is even.
    \item $a$, $b$ and $c$ are even.
    \item $\beta_{abc} = \beta_{bca} = \beta_{cab}$.
    \item $\alpha_{hkl} = -(-1)^{\frac{h+k+l}{2}}\alpha_{khl}$.
    \item $\beta_{abc} = -(-1)^{\frac{a+b+c}{2}}\beta_{bac}$.
\end{itemize}

For both the FCC and the  FCC$_{^1\!/_2}$ phases, the following coefficients are allowed
\begin{itemize}
    \item $h$ is odd, $k$ and $l$ are even.
    \item $a$, $b$, $c$ are all odd.
\end{itemize}
The FCC phase permits, in addition
\begin{itemize}
    \item $h$ is even, $k$ and $l$ are odd.
    \item $a$, $b$, $c$ are all even.
\end{itemize}

As we can see from these constraints, the FCC and FCC$_{^1\!/_2}$ phases share many Fourier coefficients. The FCC phase, however, has additional coefficients which are set to zero in the half-skyrmion phase. FCC$_{^1\!/_2}$ solutions are, therefore, at the same time particular FCC solutions. This implies that the ground state energy per baryon number of a FCC$_{^1\!/_2}$ solution can never be smaller than the  
ground state energy per baryon number of a FCC solution. The standard Skyrme model $\mathcal{L}_{\rm Sk}$ is compatible with the FCC$_{^1\!/_2}$ symmetries (it respects the symmetry 
$\sigma \to - \sigma$), so it allows both for a FCC$_{^1\!/_2}$ ground state which is symmetric w.r.t.  $\sigma \to - \sigma$ and for a FCC ground state with a spontaneously broken symmetry. It turns out that for sufficiently large $L$ the FCC ground state is realised, whereas the system settles in the more symmetric FCC$_{^1\!/_2}$ ground state at higher densities. The two phases are separated by a second order phase transition at a certain critical $L_{\rm PT}$, where the additional coefficients allowed by FCC approach zero. The pion mass term, on the other hand, is not compatible with the symmetry 
$\sigma \to - \sigma$, therefore the system is always in the FCC phase. At large densities, however, the pion mass term becomes irrelevant and the additional 
 non-zero coefficients of the FCC phase are suppressed in the limit of small $L$ \cite{Vento:2017ypn}.

\subsection{Numerical procedure}
\label{Sec:Numproc}
In order to solve the optimization problem explained above, we have to fix the value of the length $\Lt$ of the unit cell, which is an input of the crystal ansatz, and then the energy is minimized varying the Fourier coefficients using a Nelder-Mead algorithm \cite{nelder1965simplex} with the GSL C++ library. Once this process has been repeated for many different lattice length values, we will obtain a curve $\Et(\Lt)$ (energy of the unit cell) for each phase.

Such a procedure constitutes an efficient solution to the problem, because higher terms in the expansions \eqref{Sigmaexpan}, \eqref{Piexpan} only give very small contributions.  We can, therefore,   safely truncate the series to a certain finite number of terms ($N_t$) and neglect higher order terms. We take $N_t$ such that we reproduce the results in \cite{kugler1989skyrmion} up to a precision of $1$\textperthousand, concretely an energy $\Et/B = 1.038$ for an FCC half-skyrmion unit cell of length $\Lt = 4.7$, for which $N_t = 32$ terms are needed in total. This last assumption is numerically checked: the first two coefficients in that symmetry are $\alpha_{100} = 0.982$ and $\beta_{111} = -1.110$, the next-to-leading order coefficients are a $5\%$ of the first and the next order is a $0.4\%$. Due to this quick convergence, even the solution of the crystal with only 2 Fourier coefficients already provides a rather good approximation.

Once the values of the curve $\Et(\Lt)$ have been obtained, we fit them with the following function
\begin{equation}
    \frac{\Et}{B} = k + k_2\Lt + \frac{k_4}{\Lt} + C_6 \frac{k_6}{\Lt^3} + C_0 k_0\Lt^3,
    \label{E_Fit}
\end{equation}
which is motivated from the scaling behaviour of the different terms that appear in the lagrangian. An interesting observation is that the contribution to the energy of each term individually can be approximately parametrized as $\Et_i(\Lt) = K_i \Lt^{3-i}$, at least for $L \lesssim L_{\rm{min}}$. Here $K_i$ is almost a constant, and $i$ is the scaling dimension of each term. Then the energy can be expressed as the sum of the individual contributions of each term in the lagrangian. This suggests that, at least in the high density regime (which is the one of interest), there is an \emph{
approximate perfect scaling} of each term. The precision of this approximation is given by the differences $K_i \neq k_i$ and $k \neq 0$. This perfect scaling property at lower densities will be useful to fit the values of the constants $f_{\pi}$ and $e$ in the next sections.

To obtain the perfect scaling parametrization, we calculate the energy for a single value of $\Lt$ and obtain the contribution of the different terms individually to extract the constants $K_i$ (we calculate the constants $K_i$ in the case $C_6 = C_0 = 1$ for simplicity.). Then, the curve $\Et(\Lt)$ can be approximated by:
\begin{equation}
    \frac{\Et_{\text{PS}}}{B} = K_2\Lt + \frac{K_4}{\Lt} + C_6\frac{K_6}{\Lt^3} + C_0 K_0\Lt^3.
    \label{Perfect_Sc}
\end{equation}

This procedure is applied in the generalized Skyrme model. However, the inclusion of the sextic and the mass terms forces us to give numerical values for the parameters even when choosing dimensionless units, as now not all the parameters can be factored out in the Lagrangian. We choose the parameter values such that we reproduce the energy density of infinite nuclear matter at saturation, which is given by (here $p$ is the pressure and $n$ is the thermodynamical, average baryon density)
\begin{equation}
    \left.\frac{E}{B}\right|_{p = 0} = 923.3 \:\text{MeV}, \hspace{3mm} n(p = 0) = n_0 = 0.16\: \text{fm}^{-3},
    \label{Infinite_Matter}
\end{equation}
and the point $p = 0$ is identified with the minimum $\Et_{\rm min}$ of the curve $\Et(\Lt)$ (see next section). 

The baryon density of the unit cell is the number of baryons per unit cell divided by its volume, 
$n=B_\text{cell}/(2L)^2$. Here an important point is that the BCC and FCC unit cells have different baryon numbers, such that the same baryon density corresponds to different lattice parameters $L$ for different phases,
\begin{equation}
   n= \frac{B_{\rm cell}^{\text{FCC}}}{8L^3_{\text{FCC}}} = \frac{B_{\rm cell}^{\text{BCC}}}{8L^3_{\text{BCC}}} \longrightarrow \frac{L_{\text{BCC}}}{2^{1/3}} = L_{\text{FCC}}.
\end{equation}
Further, if we want to compare the $(E/B)(L)$ curves of different phases, these comparisons should be done for the same baryon density. We shall, therefore, assume that $L=L_\text{FCC}=L_\text{BCC}/\sqrt[3]{2}$ whenever such a comparison is made like, e.g., in Fig. \ref{Figure.EvsL} below.

To satisfy conditions \eqref{Infinite_Matter} at the minimum, we have to find the correct values for the physical constants, and this process must be repeated iteratively until a reasonable convergence is reached. The value of the pion mass will be fixed to its physical value $m_{\pi} = 140$ MeV. We take the value of $\lambda^2 = 7$ MeV/fm$^{3}$ motivated by the $\omega$ meson coupling \cite{Adam:2020aza}, then this coupling constant is not varied in the iteration procedure. On the other hand, the values of $f_{\pi}$ and $e$ are given as initial seeds. Since the initial values of $f_{\pi}$ and $e$ will not reproduce \eqref{Infinite_Matter}, they must be varied until we match this condition.

This iterative process of fitting the constants in the generalized model hugely increases the time of computation, since the curve $\Et(\Lt)$ must be reproduced to find the minimum each time that $f_{\pi}$ and $e$ are changed. To avoid this computational cost, we can now take advantage of the (approximate) perfect scaling property of the curve $\Et(\Lt)$  near the minimum to solve this problem much faster. In this approximation, the constants $K_i$ are already known, and only $C_6(f_{\pi}, e)$ and $C_0(f_{\pi}, e)$ will change. This approach is much faster and reproduces \eqref{Infinite_Matter} with a sufficient accuracy of a few percent. Therefore, this approximation is used to fit the values of $f_{\pi}$ and $e$ at the minimum. We just have to find the phase of minimum energy for each choice of the lagrangian, then we only need the constants $K_i$ for that specific phase.

\subsection{Results}
The values of the physical constants resulting from the perfect scaling are given in \cref{Table.Constants}. 
\begin{table}[h!]
	\centering
		\begin{tabular}{|c|c|c|c|}
			\hline
			$f_{\pi}$ (MeV) & $e$ & $\lambda^2$ (MeV/fm$^3$) & $m_{\pi}$ (MeV) \\ \hline
			137.83 & 4.59 & 0 & 0 \\ \hline
			118.83 & 4.32 & 0 & 140 \\ \hline
			160.32 & 8.59 & 7 & 0 \\ \hline
			136.85 & 6.46 & 7 & 140 \\ \hline
		\end{tabular}
		\caption{\small Values of the parameters that fit the infinite nuclear matter for each model.}
		\label{Table.Constants}
\end{table}

We show the curves $\Et/B$ for the different symmetries and for different models in \cref{Figure.EvsL}. The left upper plot (model $\mathcal{L}_{24} \equiv \mathcal{L}_{\rm Sk}$) reproduces the known results described at the beginning of this Section. A more detailed discussion of the remaining plots will be given below, where we describe the resulting phases of skyrmionic matter at different densities. In \cref{Figure.EvsL} we also use the fact that for all models except for the simplest model $\mathcal{L}_{\rm Sk}$ there exist topological energy bounds \cite{Adam:2013tga} which are tighter than the Skyrme-Faddeev bound.  We plot these topological energy bounds for each model. Although the crystals do not reach the bound, they are very close to it at the minimum. We show the values of these bounds and how far the crystals are above it in \cref{Table.Bounds} (both the bounds and the plots in \cref{Figure.EvsL} are given for the values of the parameters specified in \cref{Table.Constants}).

Further, we find that the half-skyrmion phases are well fitted to the proposed parametrization (\ref{E_Fit}) even for $L \geq L_{\min}$. However, this parametrization breaks down for large $L$ for the FCC phase, and a more complicated behaviour is observed in this region. Indeed, $\langle \sigma \rangle$ does not vanish for large $L$ in the FCC phase, but has a non-trivial dependence on $L$ which could be fitted to a hyperbolic tangent. However, for small $L$ the FCC phase is either exactly equal to the FCC$_{^1\!/_2}$ phase (a phase transition occurs) or very close to it.  In particular, the region where the FCC phase differs significantly from the FCC$_{^1\!/_2}$  phase is always beyond the minimum, i.e., for $L>L_{\rm min}$. As we shall argue below, in this region the FCC crystal is not relevant for the nuclear EoS. We will, therefore, ignore this problem and use the parameters of the fit $k_i$ that reproduce the half-skyrmion curves, which are given in Table \ref{Table.Fits}. The fit for the FCC$_{^1\!/_2}$ phase serves as a good approximation for the FCC phase in the small $L$ region.

\begin{table}[]
	\centering
		\begin{tabular}{|c|c|c|}
			\hline
			Model & Bound & Crystal value ($\%$) \\ \hline
			$\lag_{24}$ & 1 & 3.7 \\ \hline
			$\lag_{240}$ & 1.07 & 5.8 \\ \hline
			$\lag_{246}$ & 1.57 & 6.2 \\ \hline
			$\lag_{2460}$ & 1.37 & 8.0 \\ \hline
		\end{tabular}
		\caption{\small The topological bound for each model, in dimensionless units and with $C_0$, $C_6$ chosen to reproduce infinite nuclear matter. In the right column we show $[(E_{\rm min} - E_{\rm bound})/E_{\rm bound}] \times 100$, i.e., the percentage deviation of the minimum crystal energy from the bound for the FCC lattice, which provides the lowest minimum.}
		\label{Table.Bounds}
\end{table}

\begin{table}[]
	\centering
		\begin{tabular}{|c|c|c|c|c|c|}
			\hline
			Model & $k$ & $k_2$ & $k_4$ & $k_6$ & $k_0$ \\ \hline
			$\lag^{\text{FCC}}_{24}$ & 0.029 & 0.11 & 2.38 & 0 & 0 \\ \hline
			$\lag^{\text{FCC}}_{240}$ & 0.005 & 0.11 & 2.40 & 0 & 0.008 \\ \hline
			$\lag^{\text{FCC}}_{246}$ & 0.31 & 0.089 & 2.56 & 0.85 & 0 \\ \hline
			$\lag^{\text{FCC}}_{2460}$ & 0.55 & 0.050 & 1.61 & 0.89 & 0.012 \\ \hline
			$\lag^{\text{BCC}}_{24}$ & 0.014 & 0.096 & 3.00 & 0 & 0 \\ \hline
			$\lag^{\text{BCC}}_{240}$ & 0.011 & 0.096 & 3.00 & 0 & 0.004 \\ \hline
			$\lag^{\text{BCC}}_{246}$ & 0.195 & 0.087 & 2.96 & 0.239 & 0 \\ \hline
			$\lag^{\text{BCC}}_{2460}$ & 0.139 & 0.084 & 3.05 & 1.68 & 0.005 \\ \hline
		\end{tabular}
		\caption{\small Fitting constants for the numerically obtained $\Et(\Lt)$ curves.}
		\label{Table.Fits}
\end{table}

Furthermore, the constants $K_i$ that result from the perfect scaling are given in \cref{Table.PS}.
\begin{table}[h!]
	\centering
		\begin{tabular}{|c|c|c|c|c|}
			\hline
			Model & $K_2$ & $K_4$ & $K_6$ & $K_0$ \\ \hline
			$\lag^{\text{FCC}_{1/2}}_{24}$ & 0.111 & 2.43 & 0 & 0 \\ \hline
			$\lag^{\text{FCC}_1}_{240}$ & 0.114 & 2.41 & 0 & 0.0082 \\ \hline
			$\lag^{\text{FCC}_{1/2}}_{246}$ & 0.111 & 2.43 & 1.24 & 0 \\ \hline
			$\lag^{\text{FCC}_1}_{2460}$ & 0.115 & 2.41 & 1.13 & 0.0079 \\ \hline
		\end{tabular}
		\caption{\small Fitting constants for perfect scaling approximated curves.}
		\label{Table.PS}
\end{table}

The values of $f_{\pi}$ and $e$ which reproduce \eqref{Infinite_Matter}  can, in fact, be calculated exactly, since the dimensionless lagrangian $\lag_{24}$ does not depend on them,
\begin{equation}
    \left( \frac{1.28}{B_{\text{cell}}} \right)^{1/3}\Lt_{\text{min}} = f_{\pi}e, \hspace{3mm} \frac{923.3}{3\pi^2} \frac{B_{\text{cell}}}{\Et_{\text{min}}} = \frac{f_{\pi}}{e},
    \label{f-pi-e-eq}
\end{equation}
where $\Lt_{\text{min}}$ and $\Et_{\text{min}}$ denote the values of the length and energy at the minimum, and $B_{\text{cell}}$ is the baryon number of the unit cell (here we have used that the volume of the unit cell is $8\Lt^3$). In the FCC$_{^1\!/_2}$  phase, $B_{\text{cell}} = 4$, and for the model $\lag_{24}$ the exact values are $f_{\pi} = 137.77$ MeV, and $e = 4.59$, which are quite close to those obtained within the perfect scaling approximation. These values are in fact similar to those obtained from fitting the hedgehog solution to the proton \cite{Ding:2007xi}.
For the other models, we do not attempt to calculate $f_\pi$ and $e$ exactly. Instead, we calculate them from the exact scaling, see \cref{Table.Constants}, and then use \eqref{f-pi-e-eq} to find $\Lt_{\text{min}}$ and $\Et_{\text{min}}$.

In \cref{Figure.EvsL} we also see that the different terms that we include in the Lagrangian have the expected impact on the energy per baryon curve. The sextic term, due to its repulsive behaviour, shifts the length of the minimum to larger values, whereas the attractive potential term has the opposite effect. From the four models that we have studied, the same qualitative behavior is observed for the different crystalline phases in those models which do not include the potential term, i.e $\mathcal{L}_{24}$ and  $\mathcal{L}_{246}$. Indeed, for these two models, the lowest energy phase corresponds to the FCC of single Skyrmions at low densities, eventually suffering a phase transition and becoming an FCC$_{1/2}$. Such a transition was already found in \cite{Lee:2003aq}.  On the other hand, for the models 
including a pion mass potential, this transition only occurs asymptotically, as the symmetries of the FCC$_{1/2}$ phase are not compatible with a nonvanishing pion mass. Also, in all models a (first order) phase transition from the FCC$_{1/2}$ to the BCC phase is observed at high densities, as we shall explain in more detail below. Such a transition is also expected by symmetry considerations \cite{goldhaber1987maximal}.

In the rest of this section we will comment on the different density regimes at which each of these different phases become relevant, and the possible existence of phase transitions between them.

\newpage
\begin{onecolumngrid}

\begin{figure*}[h]
   \centering
    %\hspace{-2cm}
    \includegraphics[width=\textwidth]{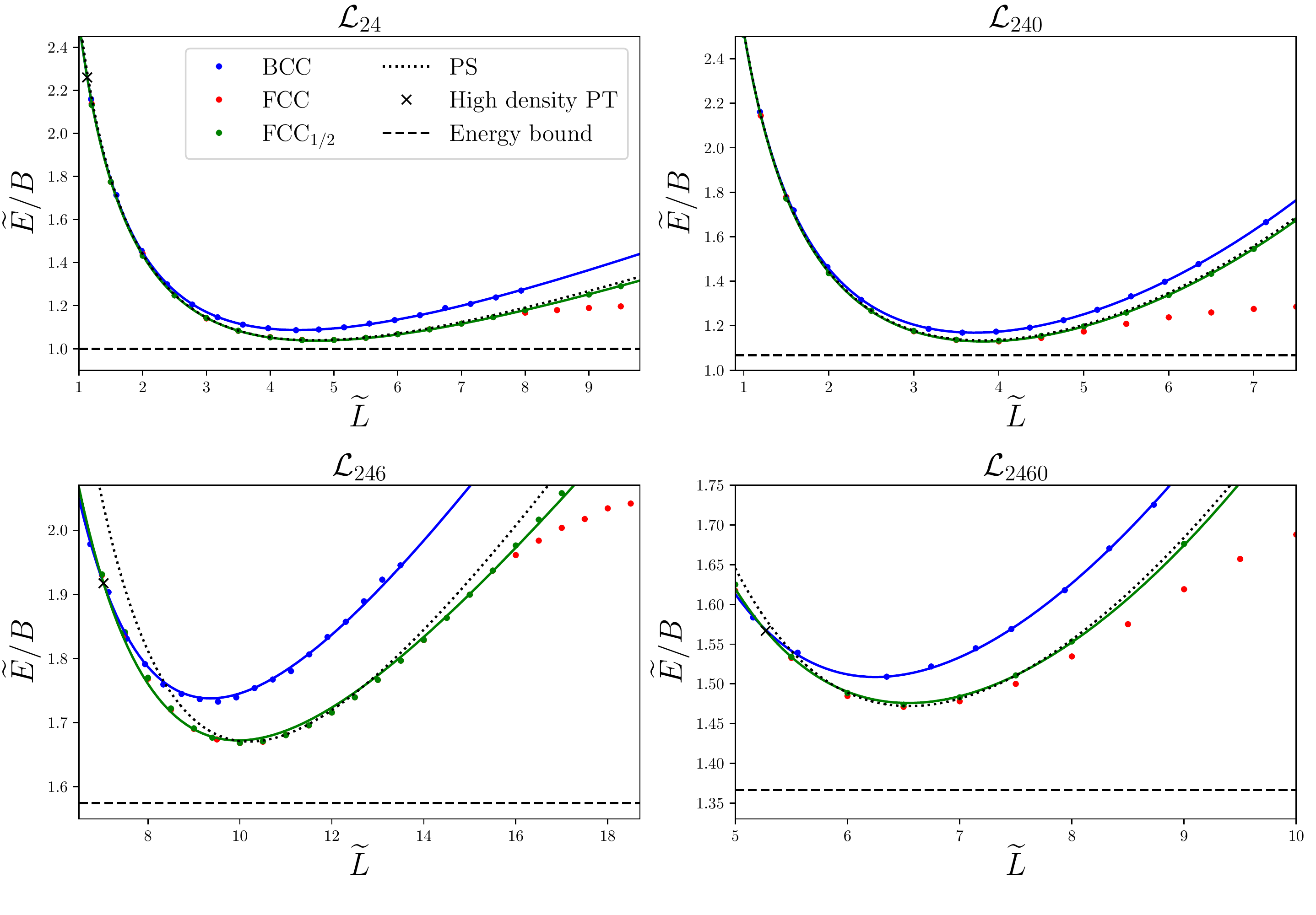}
    \caption{Energy per baryon as a function of the lattice length parameter for the four different models considered in this paper. Here, the energy per baryon is plotted in the dimensionless units of Table \ref{Table.Bounds}. Further, remember that $L=L_\text{FCC}=L_\text{BCC}/\sqrt[3]{2}$.}
    \label{Figure.EvsL}
\end{figure*}
\end{onecolumngrid}
\twocolumngrid

%\newpage

\subsection{The high density phase: transition to a fluid-like configuration}

As stated above, the BPS model shares the properties of a perfect fluid \cite{adam2010bps}. The inclusion of the sextic term in the Skyrme lagrangian, therefore, will have the effect of homogeneizing the densities in the unit cell of a crystal configuration, at least in the density regime  where the contribution from this term to the energy becomes relevant. A measure of this homogeneity may be obtained by comparing the energy density and its mean value over the unit cell. Since the sextic term scales as $1/L^3$, we expect that at the minimum of energy the density still approaches that of the FCC$_{^1\!/_2}$ crystal, whereas for decreasing values of $L$ a more homogeneous energy density (fluid behaviour) will appear, i.e., the field configuration will become more similar to a perfect fluid with homogeneous energy density.

As a measure for this effect, we define the radial energy profile (REP), i.e. the energy enclosed by a sphere of radius $r$,
\begin{equation}
    E(r) = \int_0^{r} d^3x \:\varepsilon ,
    \label{EnEnclosed}
\end{equation}
where $\varepsilon$ is the energy density (the integrand in \eqref{Energy}). For this concrete calculation, we only consider the BCC phase, because {\em i)} this is the relevant phase for high densities, and {\em ii)} the effect of homogeneization is stronger for this phase. Further, we use the smaller "unit cell" of size $L$ (because the energy density has periodicity $L$), surrounded by vacuum.   The resulting REP will grow with the radius until $r = \sqrt{3}\:L$ and take a constant value equal to the energy of the unit cell for $r \ge \sqrt{3}\:L$. In the case of a fluid, $\varepsilon_{\text{fluid}}$ is a constant, therefore we also compute the REP \eqref{EnEnclosed} for a unit cell of the same energy but with a constant energy density. The ratio $\chi$ between these two radial energy profiles tells us how far we are from the fluid-like behaviour. 
\begin{figure}[h!]
    \centering
    \includegraphics[scale=0.55]{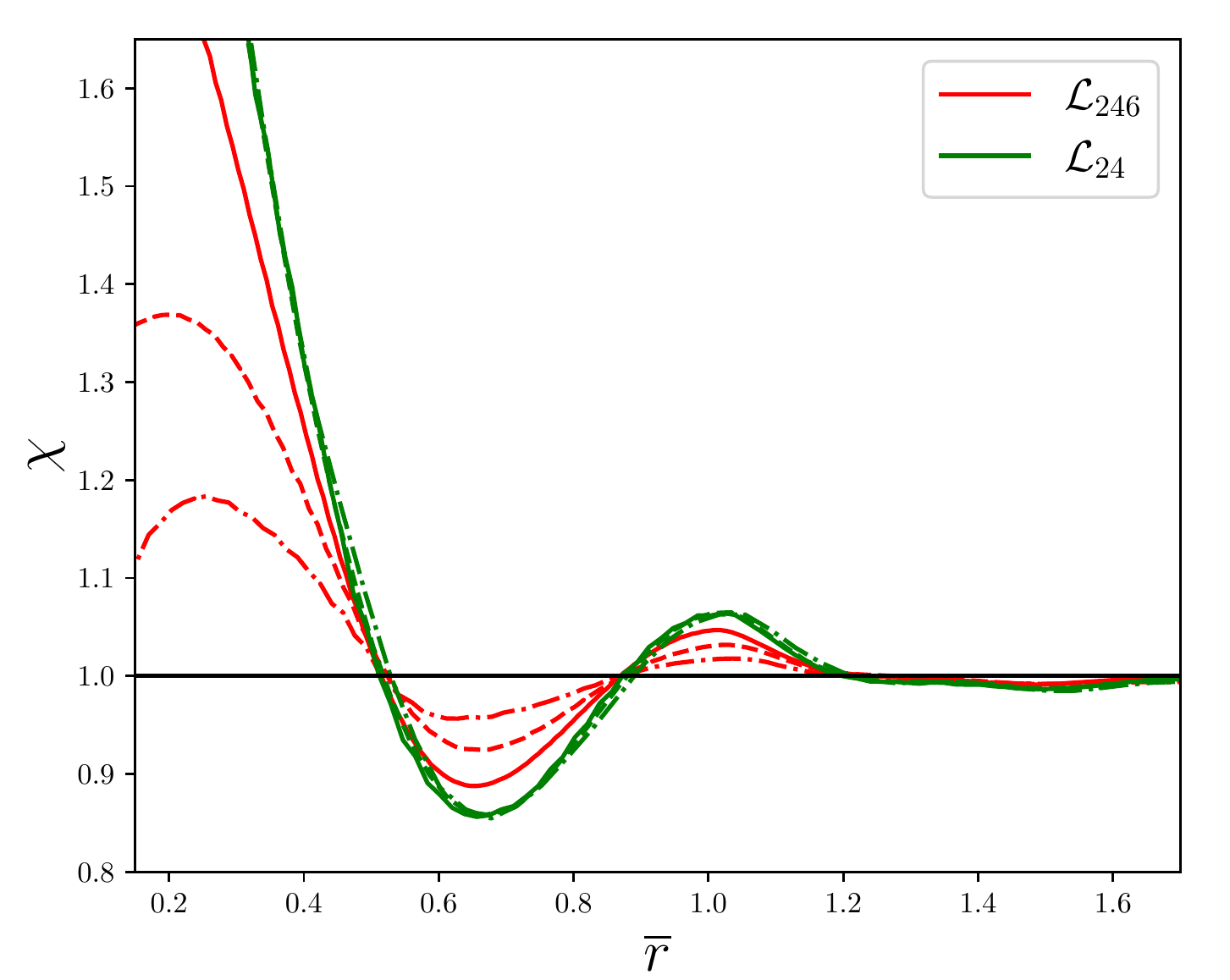}
    \caption{\small Influence of the sextic term on the ratio between the REPs for the crystal and the fluid, at $L_{\rm min}$ (solid), $\tfrac{2}{3}L_{\rm min}$ (dashed), and $\tfrac{1}{2}L_{\rm min}$ (dotline). The radial coordinate is rescaled by the lattice length $\bar{r}=r/L$. }
    \label{Figure.Fluid}
\end{figure}

\begin{figure}[h!]
        \centering
        \includegraphics[scale=0.3]{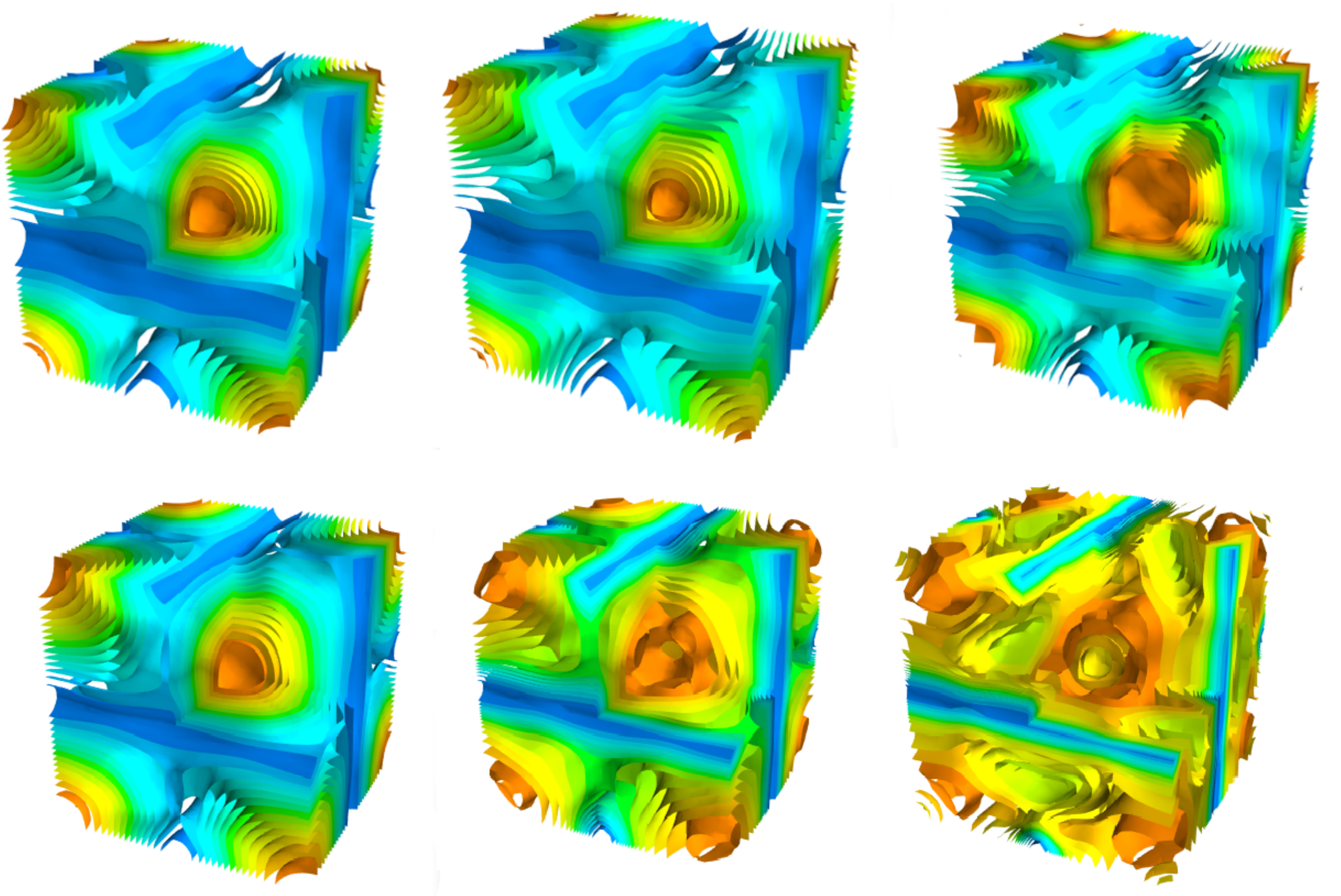}
        \caption{Evolution of the energy density of a unit cell in the BCC half-skyrmion phase with (lower row) and without (upper row) sextic term for $L=L_{\rm min},\,\, \tfrac{2}{3}L_{\rm min}$ and $\tfrac{1}{2}L_{\rm min}$}.
        \label{Figure.EDFCC}
    \end{figure}

In \cref{Figure.Fluid} we can see that the homogeneity of the energy density strongly increases with density, i.e. with decreasing values of the lattice parameter, when the sextic term is included. For the model $\mathcal{L}_{24}$ without the sextic term, on the other hand, the ratio $\chi$ between the lattice and the fluid REPs is almost independent of the density and strongly deviates from unity. In other words, skyrmionic matter remains in a crystalline phase up to very high densities without the sextic term, whereas it approaches a fluid phase when the sextic term is included. We remark that the pion mass term is irrelevant for these high-density effects.
Our findings are further illustrated by the three-dimensional energy density plots in \cref{Figure.EDFCC}. There it can be seen that regions of small energy density are almost completely expelled from the unit cell for small $L$ (high density) if the sextic term is included, leading to a much more homogeneous energy density. Without the sextic term, on the other hand, the relative sizes of the regions of small and large energy density remain almost unchanged when $L$ is varied.  

In addition, the inclusion of the sextic term is known to result in a much stiffer EoS for Skyrmionic matter at high densities \cite{Adam:2015lra}. This is one of the successes of the generalized model, as argued in \cite{Adam:2020yfv}, since it allows for more realistic neutron star maximum masses than the standard Skyrme model.

    \hspace{1cm}

In fact, one of our main results in this paper consists in the numerical confirmation of the hypothesis made in \cite{Adam:2020yfv} about the smooth transition between a pure Skyrmion crystal and a perfect fluid phase at higher densities, which is crucial to be able to describe the most massive $(M\sim 2.3 M_\odot)$ neutron stars within the (generalized) Skyrme model. We have identified the two principal factors that provide this effect, namely, the inclusion of the sextic term in the generalized model, whose repulsive character tends to homogeneize the energy density, and the transition from the FCC to BCC half Skyrmion phase (see below), which actually accelerates this proccess.

\subsection{The medium density phases and phase transitions}
\subsubsection{FCC to FCC$_{^1\!/_2}$ phase transition}

In \cref{Figure.EvsL}  we see that the effect of a nonvanishing pion mass potential has a big qualitative effect on the behavior of the $E(L)$ curve of the Skyrmion crystal at low densities. Indeed, without potential term the FCC and FCC$_{1/2}$ curves join around $L_{\rm PT} = 7.7$ and $L_{\rm PT} = 15.5$ with and without sextic term, respectively, and they have the same energy from there on. In other words, a second order phase transition from FCC to FCC$_{^1\!/_2}$ occurs at these values of the lattice parameter $L$. When we include the potential term, on the other hand, this joining never occurs exactly since the FCC$_{^1\!/_2}$ symmetries are not respected by the lagrangian. The FCC curve approaches  the FCC$_{^1\!/_2}$ curve in the chiral limit, when $\langle \sigma \rangle \rightarrow 0$. But even in this case, the two curves are essentially indistinguishable for $L\le L_{\rm min}$.

 The phase transition in the chirally symmetric case (without pion mass potential) has been previously reported in the literature \cite{Lee:2003aq}, and the vanishing of the mean value of the $\sigma$ field has been proposed as an order parameter signaling this transition, since it vanishes in the half-skyrmion crystal due to the symmetry properties of the unit cell in this phase. The physical significance of such a transition has also been extensively studied \cite{Park:2008xm,Vento:2017ypn}.  Moreover, this transition, which involves a \emph{topology change} ---in the sense that  the $4$ skyrmions of a unit cell must split into $8$ half-skyrmions with the same total baryon number--- \cite{Harada:2016tkf} has been argued to be a genuine prediction of the Skyrme crystal model for dense nuclear matter, and to have nontrivial observational effects in the EOS of neutron stars.

In these investigations, the Skyrme model (and Skyrme crystal) is typically embedded into a larger effective model motivated by QCD, containing, e.g., the dilaton field in order to recover the scale invariance of Yang-Mills theory at high density. Here we want to argue, however, that at least for the pure Skyrme model without these additional DoF, the physical relevance of this phase transition is questionable. Firstly, this phase transition always occurs at an $L_{\rm PT} > L_{\rm min}$, i.e., in a region where the energy per baryon $E/B$ {\em grows} with $L$. But this corresponds to a {\em thermodynamically unstable} region with negative pressure, as was already pointed out in \cite{kugler1989skyrmion}.

Secondly, in the next section we will show that there exists another skyrmion lattice phase with lower energy per baryon than the FCC crystal of skyrmions in the region $L\ge L_{\rm min}$. Further, this phase evolves naturally towards a half-skyrmion phase without involving any sort of change in the topology of the field configurations. Concretely, this phase describes a cubic lattice of $B=4$ skyrmions, i.e., $\alpha$ particles. In this phase, the individual $\alpha$ particles are free to occupy their preferred volume within the $B=4$ unit cell, and we find that, indeed, for large $L$ they only occupy a small fraction of the total volume.
This is in accordance with the physical picture that at low densities nuclear matter clusters into larger substructures (nuclei) and not just individual nucleons. $\alpha$ particles are good candidates for these substructures, because they are strongly bound, both in nature and in the Skyrme model.

\subsubsection{FCC$_{^1\!/_2}$ to BCC phase transition}
The energies per baryon number of the FCC and BCC phases have been compared in \cref{Figure.EvsL}. 
We appreciate in the plots of this figure that the intersections of the BCC and FCC curves, marked by a cross in all cases, always occur for rather small values of $L$ and, in particular, always for $L<L_{\rm  min}$. In this region, the FCC and FCC$_{^1\!/_2}$ curves are indistinguishable, and we use the numerical FCC$_{^1\!/_2}$ results for our calculations.
In order to obtain the correct ground state of the crystal, we have to compare the energies per baryon number at the same baryon density. This implies $L=L_\text{FCC} = L_\text{BCC}/\sqrt[3]{2}$, as explained above. From  \cref{Figure.EvsL}  we find that the minimum of the energy is always given by the FCC phase, but then at some $\Lt =\Lt_{\rm T}$ the $E/B$ curves for the FCC and the BCC phases intersect. 
The two curves have different slopes at their crossing, which implies that the phase transition is of first order and must be treated by a Maxwell construction, where the two phases are connected by a region of phase coexistence at constant pressure \cite{kugler1989skyrmion}. This implies that both the baryon density and the energy density suffer a sudden jump in the phase coexistence region when expressed as functions of the pressure. 

The values $\Lt_{\text{T}}$ at which the intersections of the two curves occur are given in Table \ref{Table.Cuts}. In the same table, we also compare $\Lt_{\text{T}}$ to $\Lt_\text{min}$ (which gives the density of nuclear matter at saturation), and we provide the pressure at the phase transition (at phase coexistence) and the jumps suffered by the energy density and the baryon density. The values of the physical coupling constants for the different models are given in Table \ref{Table.Constants}.
\begin{table}[h!]
	\centering
		\begin{tabular}{|c|c|c|c|c|}
			\hline
			\diagbox[innerwidth=5.3em, height=\line]{}{}\hspace{0.4cm}\diagbox[innerwidth=-7em, height=\line]{}{}\hspace{-0.41cm} & $\lag_{24}$ & $\lag_{240}$ & $\lag_{246}$ & $\lag_{2460}$ \\ \hline
			$\Lt_{\text{T}}$ & 1.13 & 0.98 & 7.03 & 5.27 \\ \hline
			$\Lt_{\text{min}}$ & 4.7 & 3.8 & 10 & 6.5 \\ \hline
			$\Lt_\text{T}/\Lt_{\rm{min}}$ & 0.24 & 0.26 & 0.72 & 0.80 \\ \hline
			$p_{\text{T}}$ (MeV/fm$^3$) & 6905 & 5833 & 108.2 & 53.7 \\ \hline
			$\Delta \rho$ (MeV/fm$^3$) & 669 & 484 & 41.9 & 25.8 \\ \hline
			$\Delta n$ (fm$^{-3}$) & 0.26 & 0.19 & 0.03 & 0.02 \\ \hline
		\end{tabular}
		\caption{\small FCC to BCC phase transition.}
		\label{Table.Cuts}
\end{table}

The transition to the BCC phase was expected, since the symmetries of this phase are mainly motivated at high densities. However, the transition can be produced at such high densities that they are unreachable inside NS. It is clearly seen that the inclusion of the sextic term decreases the density at which the transition is produced, making it more likely that this phase transition can occur inside NS.

%Once we have obtained the coefficients of the expansions \eqref{Sigmaexpan} that minimize the energy we can plot the baryon densities in the unit cell to compare the different shapes (Here we also include the FCC single skyrmion phase because it has a different structure at low densities).

\subsection{The low density phase: a lattice of $B=4$ skyrmions}
%Since the Skyrme model is an EFT it is only valid within a regime of densities. It is obvious that at sufficiently high densities quark matter deconfines, then the theory we have to use is QCD, this is believed to occur at densities around 40$n_0$ \cite{annala2020evidence}. 

The energy per baryon of the Skyrme crystal ansatz is bounded from below, by construction. In fact, there is a topological bound which must be satisfied at any length scale.  Furthermore, although $E/B$ gets rather close to the topological bound at the minimum --- as can be seen in \cref{Figure.EvsL}--- which corresponds to the nuclear saturation density $n_0$ (see next section), the fact that the energy per baryon grows with $L$ for $n<n_0$ shows that this particular ansatz is not valid for densities lower than $n_0$. It was first argued in \cite{Park:2002ie} that the correct minimum energy phase in this regime should correspond to an inhomogeneous phase in which skyrmions collapse into lumps with most of the space filled with vacuum. Further, in \cite{Park:2019bmi} a concrete realization of such a phase was proposed, constructed in terms of planar structures from the Atiyah-Manton instanton ansatz \cite{ATIYAH1989438}. However, this phase lacks the isotropy symmetry that one would expect from infinite nuclear matter. We now argue that there is an even simpler phase which may play the role of such an inhomogeneous phase while keeping the cubic symmetry of the unit cell, namely, the $\alpha$-particle lattice phase.

The key point is that when the parameter $L$ grows, the distance between half-skyrmions uniformly increases, and so does the unit cell as well as its energy. Nevertheless, we may assume that for distances larger than that of minimum energy, it is more energetically favourable for each unit cell to collapse into a lump with the same baryon charge, so that the Skyrme crystal fragments into a lattice of well defined $B=4$ skyrmions (one per unit cell) interacting with their neighbours. This fact was actually reported in \cite{SilvaLobo:2010acs} for the standard Skyrme model. Once the field has reached this phase, the length scale of each unit cell will not be given by the size of each skyrmion anymore, but by the distance between them, so that decreasing the density will not necessarily imply a change in the skyrmion size.

A simple way to take into account the effect of finite density is to consider skyrmions on the 3-torus (i.e. imposing periodic boundary conditions). In a first approximation, we will describe the low energy skyrmion lattice using skyrmions that preserve the cubic symmetry of the unit cell (i.e. symmetries \eqref{A1}, \eqref{A2}), the simplest of them being the cubic $B=4$ skyrmion (the $\alpha$ particle).  We then numerically obtain the energy per baryon number of alpha particles in the three-torus as a function of the torus size parameter $L$, where now $2L$ represents the distance between nearest-neighbour unit cells of the physical skyrmion lattice. We have done this calculation with the help of a gradient flow algorithm for energy minimization, on a cubic grid with $n^3$ points, with $n=2L/\delta+5$, $\delta$ being the distance between points in the grid. The extra points are needed due to the periodic boundary conditions, which were imposed by identifying the first and last two points of the grid in each dimension. The initial condition for the fields was generated from the $B=4$ rational map ansatz \cite{manton2004topological}, but using a rescaled radial coordinate for the profile function, of the form $f(r) = \tfrac{\pi}{1+(\alpha r/L)^2}$,  to account for the squeezing of the cell. The constant $\alpha$ is freely chosen so that the initial ansatz is well behaved within the unit cell. In our case, it is sufficient to take $\alpha = 5$. Once the initial condition was obtained, we run the gradient flow algorithm until an error of $\sim 10^{-4}$ in the baryon number and a convergence up to the same order in the total energy was reached. 

In our numerical calculations, we only consider the model $\mathcal{L}_{240}$ without the sextic term. 
The technical reason is that the calculations with the sextic term included become much more involved. 
The physical reason is that, at low densities, where the $\alpha$ particle lattice is relevant, the contributions of the sextic term are small and should not qualitatively change our results. More precisely, in the interior of the individual $\alpha$ particles, the sextic term will have a certain influcence, essentially consisting in the expulsion of low energy density regions. The symmetries of the $\alpha$ particles should remain unchanged, because the sextic term is invariant under volume-preserving shape changes (diffeomorphisms). In the large near-vacuum regions between the $\alpha$ particles, on the other hand, the sextic term can be safely neglected. 
\begin{figure*}
    \centering
    \includegraphics[scale=0.5]{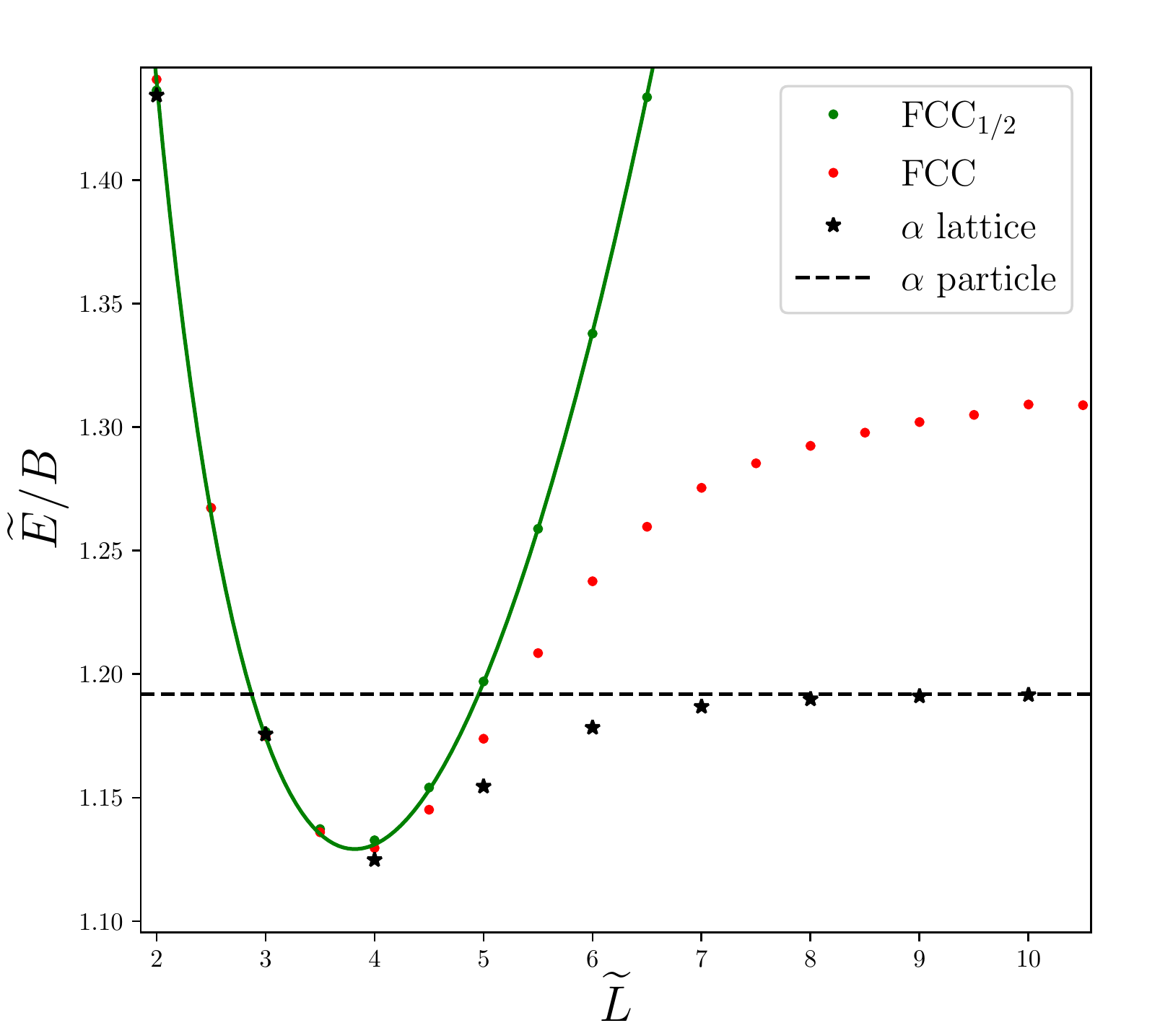}
    \caption{$E(L)$ curve for the Skyrmion crystal phases and the $\alpha$ particle lattice (asterisks)}
    \label{fig:Alphalattice_EvL}
\end{figure*}

We emphasize that the symmetries \eqref{A1}, \eqref{A2} of the $\alpha$ particle lattice are a subset of the symmetries of all the crystals which we considered. That is to say, the constraints imposed on the $\alpha$ lattice field configurations form a subset on the constraints imposed on all other lattices. This implies that the energy-per-baryon curve of the $\alpha$ lattice is bounded from above by all the other $E/B$ curves, i.e., it is a better approximation to the true minimum energy configuration that the crystals. The physical expectation is that the $\alpha$ particle lattice will lead to a strictly lower energy for sufficiently low densities (large $L$), whereas the more symmetric crystals will be recovered in the high-density region, either asymptotically or via a second-order phase transition.

\begin{figure}
\hspace{2.0cm}
\centering
    %\includegraphics[scale=0.55]{Alpha2fcc.pdf}
    %\scalebox{.4}{\input{Transition2.pgf}}
     \includegraphics[scale=0.30]{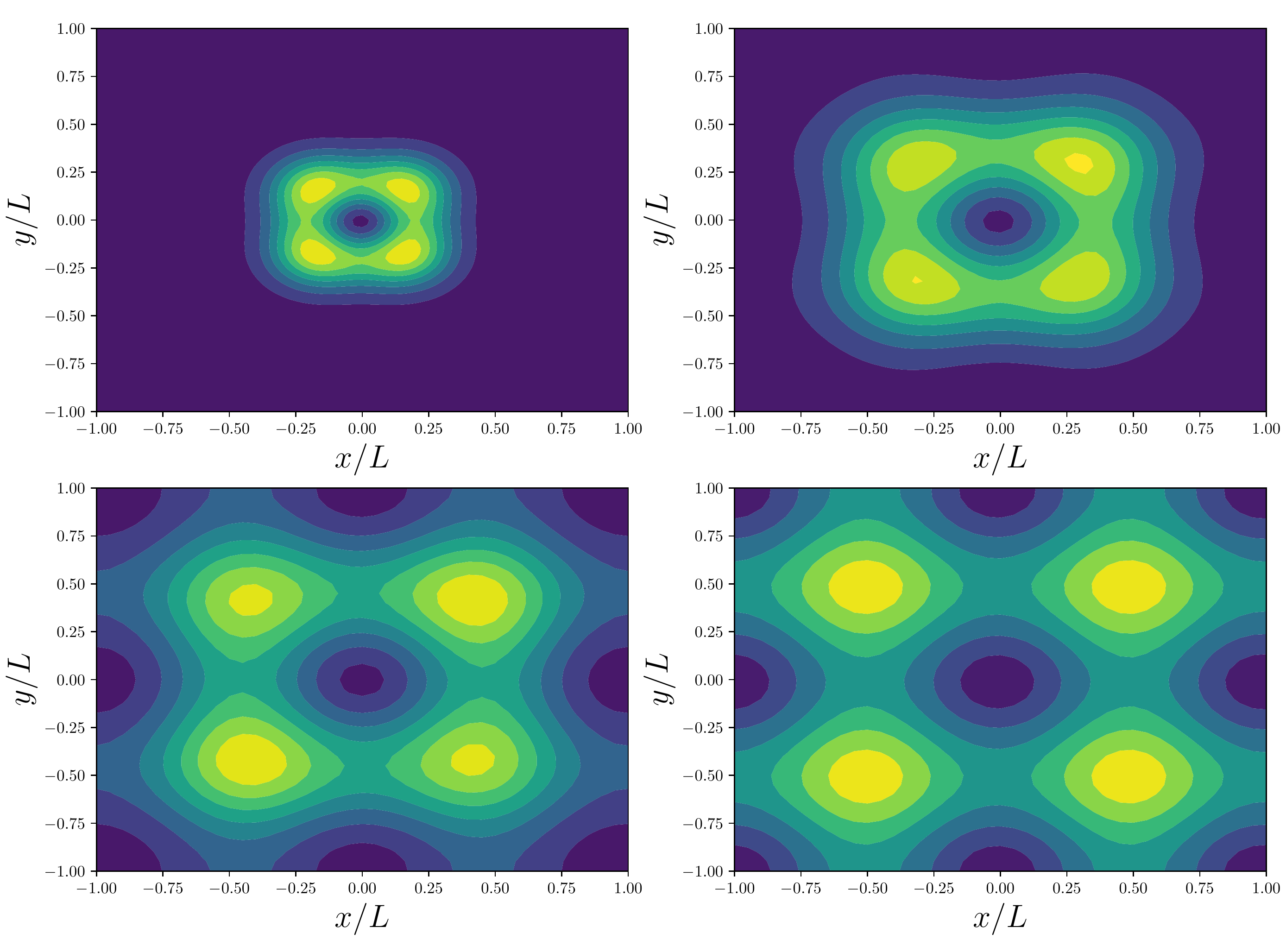}
    \caption{Energy density contours for the minimum energy field configuration in $T^3$ for different values of the torus length ($\Tilde{L}=8,5,4,2$). The color scheme is as in \cref{Figure.D}}
    \label{fig:alpha_transition}
\end{figure}

In \cref{fig:Alphalattice_EvL} we can see that the energy per baryon number of $\alpha$ particles on $T^3$ tends to the isolated $B = 4$ Skyrmion at low densities, and that the $\alpha$-lattice phase has less energy per baryon than all the skyrmion crystal phases for $L>L_{\rm min}$, so that the former is a strictly better ansatz for the low density region than the rest. Indeed, our numerical results indicate that there is a transition near the minimum of energy, such that the interpolated curve between the two phases (FCC crystal before $L_{\rm min}$ and $\alpha$-lattice just after) describes the correct behavior of the skyrmion matter in this range of densities. The energy density plots of \cref{fig:alpha_transition} confirm the behaviour described above. For sufficiently large $L$, the $\alpha$ particle only occupies a small fraction of the unit cell. For small $L$, instead, we recover the half-skyrmion structure of the FCC$_{^1\!/_2}$ lattice. To appreciate that the energy density of \cref{fig:alpha_transition} approaches that of \cref{Figure.D} in the limit of small $L$, we have to shift the energy density plot of \cref{Figure.D}, left panel, by $L/2$ in the $x$ and $y$ directions. The reason is that in \cref{fig:alpha_transition} the $\alpha$ particle is always placed in the center of the unit cell, whereas in \cref{Figure.D} the half-skyrmions are placed at the corners of the unit cell. We would like to remark that the transition from the $\alpha$ lattice to the FCC$_{^1\!/_2}$ lattice happens quite naturally, and no topology change occurs, since the half-skyrmion structure of the energy density is already present in the structure of the $\alpha$ particles, as can be seen in \cref{fig:alpha_transition}. 

%\begin{figure*}
%\hspace{0.5cm}
%    \scalebox{.33}{\input{Transition_2.pdf}}
%    \caption{Energy density contours for the minimum energy field configuration in $T^3$ for different values of the torus length ($\Tilde{L}=8,5,4,2$). The color scheme is as in \cref{Figure.D}}
%    \label{fig:alpha_transition}
%\end{figure*}

Indeed, it is well-known that already the single $B=4$ skyrmion first reported in \cite{Braaten:1989rg}, corresponding to the $L\to \infty$ limit of our $\alpha$ particles on $T^3$, shows this half-skyrmion substructure, see e.g., \cite{Manton:2011mi}. We emphasize that this half-skyrmion substructure of the $\alpha$ particles is not imposed in the numerical procedure. Instead, it is a property of the resulting solution.

Furthermore, this transition to the $\alpha$ particle lattice renders the difference between the minimum energy and the energy at $L\rightarrow \infty$ not only finite, but very small (about $\sim 5\%$). Obviously, we could improve the minimization of this difference even further by considering a larger unit cell containing, e.g., the $B = 32$ or $B = 108$ solutions, which have a slightly lower energy per baryon than the alpha particle and share its cubic symmetry. However, this difference would be rather small, of the order of $1$ or $2$ percent, so the $\alpha$-lattice approximation, being significantly simpler from the numerical point of view, already constitutes a good candidate for the description of skyrmion matter at low densities.

\section{The Skyrme crystal EoS}

Before discussing the EoS resulting from the crystals which minimize the energy per baryon in the different density regions, it is useful for our purposes to show a figure similar to \cref{Figure.EvsL}, but where the baryon density $n$ is used as the independent variable (horizontal axis) and the $E/B$ vs. $n$ curves are shown in physical units. That is to say, the true minimum (the minimum of the FCC or FCC$_{^1\!/_2}$ curve) is located at the saturation density $n_0$ and takes the value $(E/B)_0 = 923.3\;$MeV.

It is clearly visible from \cref{Figure.Evsn} that the increase in energy per baryon with $n$ is much steeper for models including the sextic term, so that much bigger energies are reached at the same baryon density. This implies a much stiffer EoS. Further, the FCC-to-BCC phase transition occurs in a region of 3-5 $n_0$ when the sextic term is included, which is clearly relevant for the interior of sufficently heavy NS. This is no longer the case without the sextic term. Also the second order phase transition from FCC to FCC$_{^1\!/_2}$ for the models without potential in the unstable region $n < n_0$ can be clearly identified. When the pion mass potential is included, this phase transition turns into an asymptotic approach.

The equation of state (EoS) $\rho(p)$ is the relation between the (thermodynamical, average) energy density $\rho$ of a system and the pressure $p$ applied to it. Both magnitudes $\rho$ and $p$ as well as the baryon density $n$ can be obtained from the crystal energy, using their thermodynamical definitions 
\begin{align}
    \rho &= \frac{E}{V} = \frac{N_{\text{cells}}\:E_{\text{cell}}}{N_{\text{cells}}\:V_{\text{cell}}} = \frac{E_{\text{cell}}}{8L^3},
    \label{eos1} \\[2mm]
    p &= -\frac{\partial E_{\text{cell}}}{\partial V_{\text{cell}}} = -\frac{1}{24L^2}\frac{\partial E_{\text{cell}}}{\partial L}, \label{eos2} \\[2mm]
    n &= \frac{B_{\text{cell}}}{V_{\text{cell}}} = \frac{B_{\text{cell}}}{8L^3}. \label{eos3}
\end{align}

Again, all these quantites remain finite in the thermodynamical limit. Since the energy of the unit cell is also a function of $L$ \eqref{E_Fit} we have to solve the equation $L(p)$ to finally obtain the EoS. 
For the standard Skyrme Lagrangian it is possible to invert this function analytically. But once the sextic and potential terms are included, this inversion must be done numerically. Further, it is obvious from the above definitions that the regions $L>L_\text{min}$ (or $n<n_0$), where $E/B$ grows with $L$, correspond to thermodynamically unstable regions of negative pressure. This remains true even if the $\alpha$-particle phase is included, although this phase ameliorates the problem. We shall exclude those regions from our plots for the EoS which are, therefore restricted to $p\ge 0$ ($n\ge n_0$). In \cite{Adam:2020yfv} the EoS was extended to $n< n_0$ by  
a smooth interpolation to the standard nuclear physics EoS 
of \cite{Sharma:2015bna}. Below we shall discuss possibilities to overcome this restriction and to derive a purely Skyrme model EoS valid for all densities.

\begin{figure}[h!]
    \centering
    \includegraphics[scale=0.6]{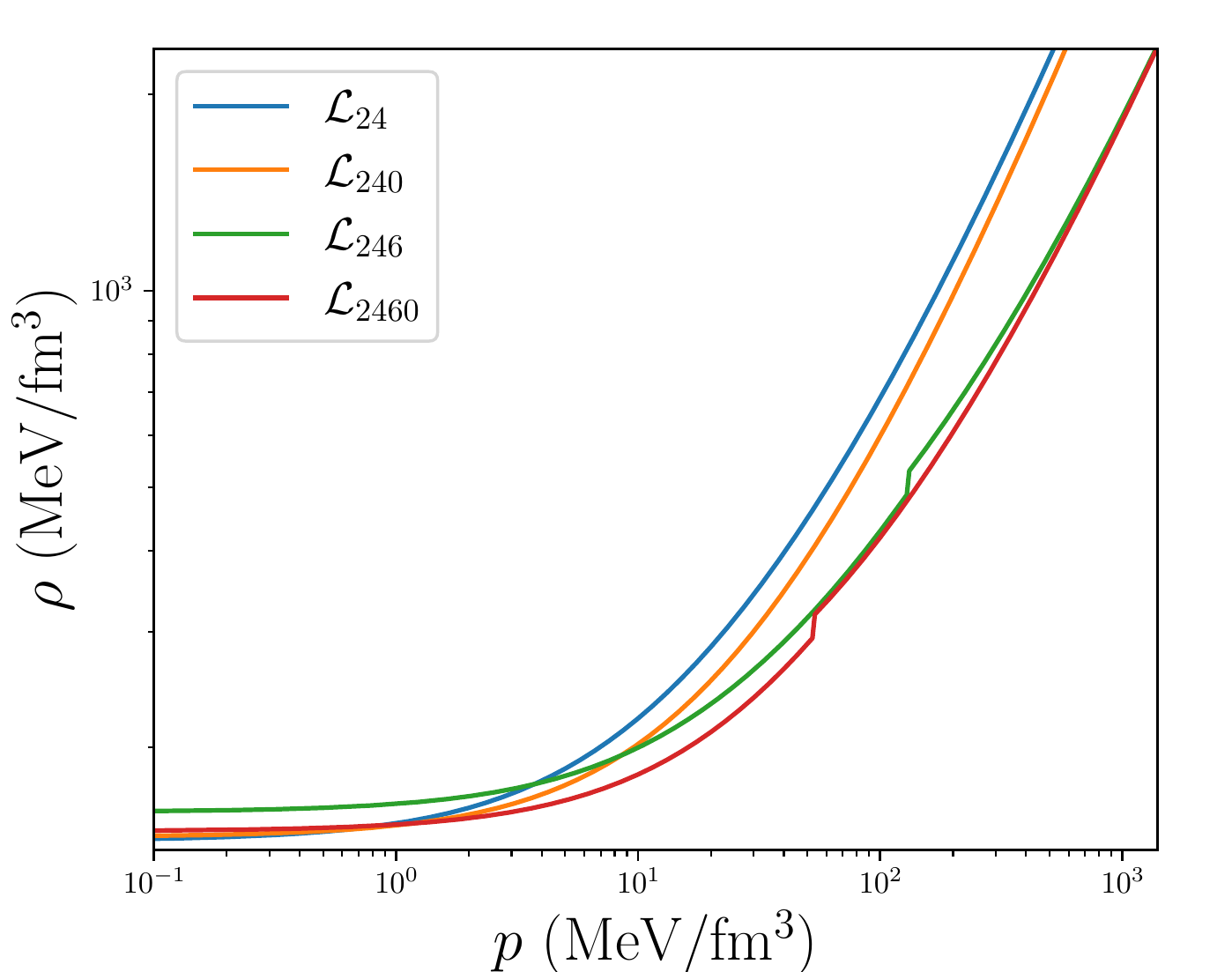}
    \caption{Equation of State for all four models}
    \label{Figure.EoS}
\end{figure}

The EoS resulting from \eqref{eos1} - \eqref{eos3} are shown in \cref{Figure.EoS}, in which the energy and baryon densities are plotted against the corresponding pressure, for a range of values which have been shown to be physically relevant for matter inside neutron stars \cite{Adam:2020aza}.

\newpage
\begin{onecolumngrid}

\begin{figure*}[htb!]
   \centering
    %\hspace{-2cm}
    \includegraphics[width=\textwidth]{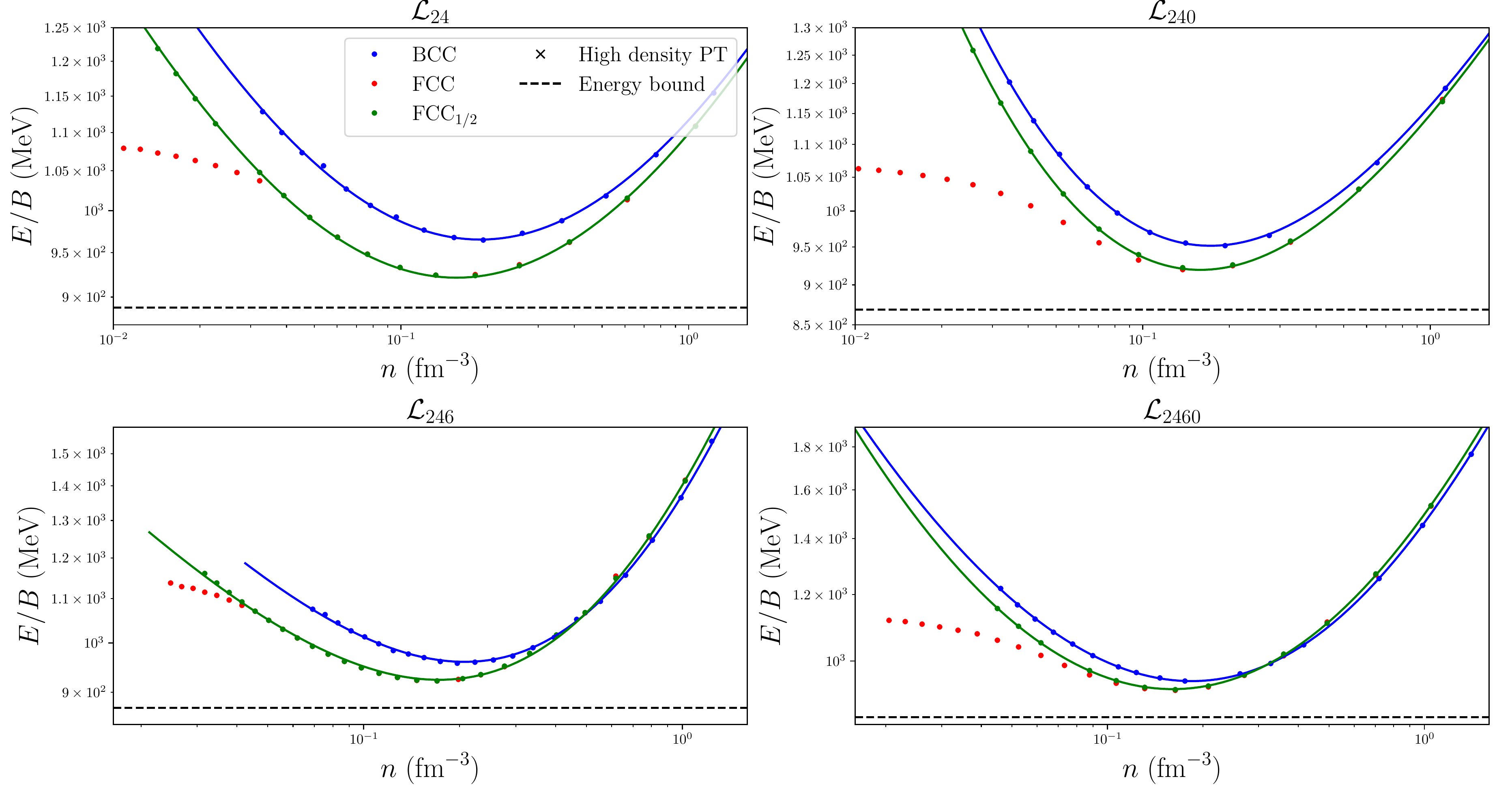}
    %\scalebox{0.43}{\input{Figures/EB4figs.pdf}}
    \caption{ Energy per baryon of the unit cell vs baryon density of the different crystals in various models. The true (FCC) minima are fitted to the energy and baryon density of symmetric, infinite nuclear matter at saturation.}
    \label{Figure.Evsn}
\end{figure*}
\end{onecolumngrid}
\twocolumngrid

As explained in \cref{Sec:Numproc}, the free parameters of each model are fitted so that the minimum energy per baryon in the crystal corresponds to that for saturated, infinite nuclear matter. In particular, this implies that all the EoS depicted in \cref{Figure.EoS} must converge to the same point in the $(\rho,p)$ plane as $p$ goes to zero. However, since we have determined the values of $f_{\pi}$ and $e$ using the perfect scaling approximation, these curves do not reproduce exactly \eqref{Infinite_Matter}. Nevertheless, the largest difference is produced for the $\lag_{246}$ case and it is about 6$\%$. 
Furthermore, as shown in the previous section, a phase transition between the FCC$_{^1\!/_2}$ and BCC phases is expected to occur in the high density region. Indeed, we take into account such a transition in the equation of state by smoothly joining the corresponding EoS of the two different phases via the Maxwell construction, i.e., the points at which $\pdv{E_\text{cell}}{V_\text{cell}}$ coincide for each phase are joined through a straight line tangent to both curves. This means that, at a certain value of the pressure, the baryon and energy densities suffer a sudden jump, which corresponds to a first order phase transition.

In \cref{Figure.EoS} we see that the inclusion of the sextic term in the Lagrangian significantly stiffens the resulting EoS, which was of course expected due to the incompressible character of matter in the $BPS$ Skyrme submodel, towards which the generalized model tends at large pressures.  Another effect of the inclusion of the sextic term is that the FCC-to-BCC first-order phase transition is shifted to smaller densities which may become relevant for the core region of heavy NS. 

\subsection{Towards a description of asymmetric nuclear matter and NS crusts within the Skyrme model}
Despite constituting only $\sim 1\%$ of the total stellar mass, the crust, defined as the external region of a neutron star with densities $\rho \lesssim 10^{14} \rm g \, cm^{-3}$, plays an important role for determining many observational properties, such as the tidal deformability or the cooling rate via neutrino emission. It is also a crucial element to explain radio pulsar glitches \cite{Chamel:2008ca}.

A good effective model for nuclear matter in neutron stars, therefore, should be able to describe matter at such (and lower) densities. However, it is precisely at these density regimes where the Skyrme model approach to nuclear matter becomes problematic, because the energy density obtained from the thermodynamical definition in all the phases studied above reaches a finite value at zero pressure, due to the presence of a minimum in the curve $E(L)$.

The presence of such a minimum in the binding energy is a feature shared by all models of symmetric nuclear matter \cite{Sammarruca:2010gc,Sammarruca:2021bpn,Ekstrom:2015rta,DBHFapproachNucmat}, and signals the point at which nuclear matter is most bounded, referred to as the nuclear saturation point in standard nuclear physics literature.
This is, in fact, the main reason why we interpret the classical Skyrme crystal configurations as models for symmetric nuclear matter and identify the minimum of $E/B$ with the nuclear saturation point. 
This minimum, however,  does not show up in physical nuclear matter, and deviations from the symmetric nuclear matter model become relevant near this point. Indeed, our approach to nuclear matter has only taken into account the \emph{classical} properties of the Skyrmion crystals. In other words, we have not taken into account, for example, the so called \emph{symmetry energy} ---which in the Skyrme model results from the quantization of isospin---, that is of great importance when describing nuclear matter at these density regimes. The correct treatment of quantum effects, such as isospin interactions due to the difference between the proton and neutron number, as well as the Coulomb forces, require a detailed analysis that will be developed in a future publication. 

It is expected \cite{Baskerville:1996he}, however, that the quantum corrections to the skyrmion crystal will only be relevant precisely in the intermediate density regime at which the $E(L)$ curve presents its minimum, and that such a minimum will disappear when these quantum effects are properly taken into account. Indeed, in \cite{Baskerville:1996he} a correction of about $4\%$ to the energy at the minimum was obtained from the isospin contribution in the half Skyrmion phase, to be compared with the $5\%$ difference in energy per baryon between the minimum and the $L\rightarrow \infty$ limit in the new $\alpha$ lattice phase. 

To summarize, there are strong indications that a more complete description of Skyrmionic matter which includes both quantum and Coulomb effects can erase the minimum in $E/B$ and, therefore, lead to an EoS that is valid also at low densities $n<n_0$. 
In this case, we would be able to construct a genuine equation of state for physical nuclear matter and neutron stars from the Skyrme model alone, valid for the full range of densities, hence able to describe both the ultradense NS cores and the solid NS crusts within a single effective model.

\section{Conclusions}
It was the main purpose of the present paper to provide a detailed investigation of the different phases of Skyrme crystals in the generalized Skyrme model defined in Eq. \eqref{Lag} and the resulting EoS, having in mind mainly its application to nuclear matter and the description of neutron stars. More concretely, we
\begin{enumerate}
    \item confirmed the importance of the sextic term in the generalized Skyrme Lagrangian for describing nuclear matter at sufficiently high densities. As conjectured in \cite{Adam:2020yfv}, the inclusion of this term leads to an EoS which behaves like a crystal for low densities but approaches a perfect fluid in the high-density limit. The sextic term is crucial to describe NS cores implying, in particular, their perfect-fluid behavior, because it allows to describe NS with masses up to $M \sim 2.3 M_\odot$ which have been observed recently.
    \item presented a clear picture of the different known crystalline phases of the Skyrme model, as well as the possible transitions between phases, and discussed whether or not they may appear as the true ground states for symmetric nuclear matter at some given density based on physical grounds. Specifically, we found that the FCC-to-BCC phase transition, which occurs at unrealistically high densities in the standard Skyrme model, is shifted to densities of 3-5 $n_0$ when the sextic term is included.  This density region is certainly relevant for the inner core of sufficiently heavy neutron stars. The FCC-to-FCC$_{^1\!/_2}$ phase transition, on the other hand, is unlikely to be of relevance for the nuclear matter EoS. First of all, it occurs in the thermodynamically unstable region $n<n_0$ of the classical Skyrme crystal. Secondly, we found that there exists another phase of a lattice of $\alpha$ particles with strictly lower energy in this region.
    \item described this new phase, the $\alpha$-particle lattice, which is obtained numerically using a gradient flow procedure, and represents (to our knowledge) the best approximation for the ground state of skyrmion matter past the minimum of the energy-per-baryon curve. Furthermore, this phase has some appealing characteristics to make it a good model for nuclear matter in neutron star crusts, which are believed to consist of well defined nuclei sparsely distributed in a lattice.
   
\end{enumerate}
In this paper, we only investigated classical Skyrme crystals which, up to a certain degree, can be viewed as models for symmetric nuclear matter. The resulting EoS could still be used for sufficiently high densities, where it gives a reasonable description, and matched to a standard nuclear physics EoS at some density $n_* > n_0$ to calculate the resulting neutron star EoS, as we did in \cite{Adam:2020yfv}. However, our ultimate objective is to achieve a good phenomenological description of the nuclear matter EoS at all regimes of density and pressure using only the Skyrme model - or some extensions thereof - to represent baryonic DoF.

A next important step in this direction would be the inclusion of both quantum corrections and Coulomb effects into our Skyrme crystal calculations. These corrections, which would, e.g., incorporate effects of the symmetry energy, may lead to a Skyrme-model based EoS which is valid for the whole density range of neutron stars, from the inner core to the crust, thus providing us with an approximate EoS for asymmetric nuclear matter. As argued in the previous section, preliminary results involving the addition of isospin quantum corrections to the Skyrmion crystal energy per baryon are very encouraging.

Another important issue is the inclusion of further degrees of freedom besides the pions. Indeed, the appearance of hyperons and, in particular, the condensation of kaons is expected to occur at sufficiently high densities in the core of heavy NS, leading to a softer EoS. Previous investigations suggest that in the standard Skyrme model, kaon condensation sets in at about 3.5 $n_0$  \cite{Westerberg:1994hu}. However, the magnitude of its effect on the resulting EoS, as well as the effect of the sextic term in the kaon condensation onset are both worth investigating. Further, the importance of vector mesons, concretely the rho meson, for the correct formation of alpha particle clusters in light nuclei has been demonstrated recently in \cite{Naya:2018kyi}. It is perfectly conceivable that the inclusion of rho mesons is also required for a realistic description of nuclear matter. All these questions require further studies.

At this point, it is interesting to recall the main differences between the Skyrme model, on the one hand, and other effective field theory approaches like ChPT, on the other. In those theories, the nucleons are treated as quantum mechanical point particles, and many resulting properties of strongly interacting matter are related to the corresponding quantum effects, like the degeneracy pressure or the in-medium formation of Cooper pairs leading to a neutron superfluid. In the Skyrme model, instead, the nucleons are extended objects already classically, and the most important question for the determination of the EoS is how these finite chunks of matter must be arranged in order to minimize the energy per baryon number. Quantum corrections can, in principle, be included in the Skyrme model description of nuclear matter, but experience tells us that they are subleading in many cases.
In other words, the Skyrme model approach to nuclear matter assumes that, at least at sufficiently high densities, the extended, non-point-like character of the nucleons is their most important property.
Physical nucleons are extended objects and, in addition, the nuclear force becomes strongly repulsive at short distances, therefore this assumption seems reasonable.

In any case, our point of view is that one should simply develop the Skyme model predictions for strongly interacting matter properties as far as possible, work out its consequences, and compare with the available data, especially those extracted from neutron star observations, which currently seem to be the most reliable ones at high densities. 
Such an open-minded approach is all the more justified because {\em i)} experimental results on strongly interacting matter above saturation density are still quite scarce and {\em ii)} more standard approaches face some difficulties in explaining several neutron star puzzles like, e.g., the rather high observed maximum NS masses, or the so-called hyperon puzzle. In addition, already a rather simple Skyrme-model based approach to neutron stars leads to very reasonable results for NS properties  \cite{Adam:2020yfv}, as mentioned above.

To summarize, we think that our results present an important next step towards the final goal of a realistic description of nuclear matter and neutron stars within the framework of the (generalized) Skyrme model. 

\begin{acknowledgements}
The authors would like to thank C. Naya for helpful discussions.

Further, the authors acknowledge financial support from the Ministry of Education, Culture, and Sports, Spain (Grant No. FPA2017-83814-P), the Xunta de Galicia (Grant No. INCITE09.296.035PR and Centro singular de investigaci\'on de Galicia accreditation 2019-2022), the Spanish Consolider-Ingenio 2010 Programme CPAN (CSD2007-00042), Maria de Maetzu Unit of Excellence MDM-2016-0692, and the European Union ERDF.
AW is supported by the Polish National Science Centre,
grant NCN 2020/39/B/ST2/01553.
AGMC is grateful to the Spanish Ministry of Science, Innovation and Universities, and the European Social Fund for the funding of his predoctoral research activity (\emph{Ayuda para contratos predoctorales para la formaci\'on de doctores} 2019). MHG is also grateful to the Xunta de Galicia (Conseller\'ia de Cultura, Educaci\'on y Universidad) for the funding of his predoctoral activity through \emph{Programa de ayudas a la etapa predoctoral} 2021.
\end{acknowledgements}

%

%\bibliography{biblio}% Produces the bibliography via BibTeX.
\end{document}